\documentclass[reqno]{amsart}
\usepackage[lite]{amsrefs}
\usepackage{mathtools,amsmath,amssymb}
\usepackage[pdfencoding=auto, psdextra]{hyperref}
\hypersetup{colorlinks=true,allcolors=[rgb]{0,0,0.6}}
\usepackage[hmargin= 1.3 in,vmargin=1 in]{geometry}
\usepackage{bbm,mathrsfs}
\usepackage{dsfont}
\usepackage{xcolor}
\usepackage[all,cmtip]{xy}
\usepackage{appendix}
\usepackage{graphicx, tikz, tikz-cd}

\newcommand{\A}{\mathcal A}
\newcommand{\Aan}{\mathcal A_{\mathrm{an}}}
\newcommand{\norm}[1]{\left\lVert #1\right\rVert}
\newcommand{\ad}{\operatorname{ad}}
\newcommand{\Schatten}{\mathfrak S}

\newcommand{\R}{\mathbb{R}}
\newcommand{\C}{\mathbb{C}}

\newcommand{\N}{\mathbb{N}}

\newcommand{\eps}{\varepsilon}

\newcommand{\e}{\mathrm{e}}
\newcommand{\dd}{\mathrm{d}}
\newcommand{\Tr}{\mathrm{Tr}}
\newcommand{\Nop}{(\mc{N}_{\eps} + 1)}

\newcommand{\M}{\mathcal{M}}

\newcommand{\mcN}{\mathcal{N}}

\newcommand{\mfh}{\mathfrak{H}}

\newcommand{\mc}{\mathcal}

\newcommand{\Real}{\operatorname{Re}}

\newcommand{\ii}{\mathrm{i}}

\numberwithin{equation}{section}

\usepackage{graphicx, color}

\theoremstyle{plain} 
\newtheorem{theorem}{Theorem}[section]
\newtheorem*{theorem*}{Theorem}
\newtheorem{lemma}[theorem]{Lemma}
\newtheorem*{lemma*}{Lemma}
\newtheorem{corollary}[theorem]{Corollary}
\newtheorem*{corollary*}{Corollary}
\newtheorem{proposition}[theorem]{Proposition}
\newtheorem*{proposition*}{Proposition}

\newtheorem*{conjecture*}{Conjecture}

\theoremstyle{definition} 
\newtheorem{definition}[theorem]{Definition}
\newtheorem*{definition*}{Definition}

\newtheorem*{example*}{Example}
\newtheorem{remark}[theorem]{Remark}
\newtheorem*{remark*}{Remark}
\newtheorem{assumption}[theorem]{Assumption}
\newtheorem*{assumption*}{Assumption}
\newtheorem{notation}[theorem]{Notation}
\newtheorem*{notation*}{Notation}

\title[Microscopic Derivation of the Classical KMS Condition for 1D Bose gases]{Microscopic Derivation of the Classical KMS Condition for One-Dimensional Bose Gases}
\author{Zied Ammari}
\address{Université Marie \& Louis Pasteur, CNRS, LmB (UMR 6623), F-25000 Besançon, France}
\email{zied.ammari@umlp.fr}
\author{Shahnaz Farhat}
\address{Department of Mathematics, University of T\"ubingen, 72076 T\"ubingen, Germany.}
\email{shahnaz.farhat@uni-tuebingen.de}
\author{S\"oren Petrat}
\address{School of Science, Constructor University Bremen, Campus Ring 1, 28759 Bremen, Germany.}
\email{spetrat@constructor.university}
\author{Andrew Rout}
\address{Dipartimento di Matematica, Politecnico di Milano, P.zza Leonardo da Vinci 32, 20133, Milano, Italy.}
\email{andrewjames.rout@polimi.it}


\begin{document}

\begin{abstract}
The Kubo--Martin--Schwinger (KMS) condition provides an intrinsic characterization of thermal equilibrium, both in quantum and classical statistical mechanics. 
In this work, we establish  stability of the KMS condition through the quantum-to-classical limit  for  one-dimensional defocusing Bose gases. More precisely, we consider scaled quantum KMS states in the (mean-field) semiclassical regime and prove that every associated Wigner measure satisfies the classical KMS condition corresponding to the nonlinear Schr\"odinger Hamiltonian. This result extends to an infinite-dimensional  setting an earlier derivation of the classical KMS condition for finite-dimensional Bose--Hubbard systems. It also provides a complementary perspective on the derivation of nonlinear Gibbs measures from many-body quantum statistical mechanics.
\end{abstract}

\maketitle

\tableofcontents


\section{Introduction}\label{sec:intro}
The Kubo--Martin--Schwinger (KMS) condition is one of the fundamental formulations of thermal equilibrium in statistical mechanics. In finite-dimensional quantum systems, equilibrium at inverse temperature $\beta>0$ is described by the Gibbs state
\begin{equation*}\label{eq:intro-gibbs}
\omega_\beta(A)
:=
\frac{\Tr\!\left(e^{-\beta H} A\right)}
{\Tr\!\left(e^{-\beta H}\right)},
\end{equation*}
where $H$ is the Hamiltonian of the system. If
\[
\alpha_t(A):=e^{\ii tH}Ae^{-\ii tH}
\]
denotes the corresponding Heisenberg dynamics, cyclicity of the trace yields the KMS identity
\begin{equation}\label{eq:intro-quantum-kms}
\omega_\beta\bigl(A\alpha_{\ii\beta}(B)\bigr)
=
\omega_\beta(BA),
\end{equation}
 where $\alpha_{\ii\beta}(\cdot)$ denotes the analytic continuation of the dynamics evaluated at the imaginary time $\ii\beta$. In finite dimensions, this condition characterizes the Gibbs state. Its real importance, however, appears in infinite-volume systems,
 where a Gibbs density matrix may no longer be available and several equilibrium states may coexist. In this setting, the KMS condition provides an intrinsic characterization of equilibrium that does not rely on an explicit density matrix.

\medskip
There is a corresponding notion of KMS equilibrium in classical statistical mechanics. Formally, its relation with the quantum condition can already be seen at the level of commutators. Introducing a semiclassical parameter $\varepsilon>0$, one may rewrite the quantum KMS relation \eqref{eq:intro-quantum-kms} as
\[
\omega_\beta\left(\frac{AB-BA}{\ii\varepsilon}\right)= \omega_\beta\left(A\,
\frac{B-\alpha_{\ii\beta}(B)}{\ii\varepsilon}\right).
\]
In the semiclassical regime, commutators are expected to converge to Poisson brackets, while the infinitesimal imaginary-time evolution is governed by a classical Hamiltonian $h$. At a purely formal level,
\[
\frac{AB-BA}{\ii\varepsilon}
\longrightarrow
\{a,b\},
\qquad
\frac{B-\alpha_{\ii\beta}(B)}{\ii\varepsilon}
\longrightarrow
\beta\{h,b\}.
\]
Thus if a family of quantum equilibrium states admits a semiclassical limit $\mu$, one expects $\mu$ to satisfy an identity of the form
\begin{equation*}\label{eq:intro-classical-kms}
\mu\left(\{a,b\}\right)
=
\beta \,\mu\left(a\,\{h,b\}\right),
\end{equation*}
for a suitable class of observables $a$ and $b$. This is the classical KMS condition introduced in the statistical-mechanical setting by Gallavotti and Verboven \cite{GV75}; see also the characterization of classical equilibrium states by Aizenman, Goldstein, Gruber, Lebowitz, and Martin \cite{AGGLM77} and \cite{AS23} for more recent development on the subject. 

\medskip
The purpose of the present work is to justify the stability of the KMS condition through the quantum-to-classical limit for an infinite-dimensional interacting bosonic system. We consider a one-dimensional defocusing Bose gas in the mean-field regime and prove that every Wigner measure associated with its quantum KMS (Gibbs) states satisfies the classical KMS condition for the corresponding nonlinear Hamiltonian.  More precisely, we prove the following  statement under the Assumptions~\ref{A_assumption} and \ref{V_assumption} (see Section \ref{sec3:Bose_gas}). 
\begin{theorem*}
Let $(\omega_{\beta,\varepsilon})_{\varepsilon\in(0,1)}$ be the family of quantum $\beta$-KMS states of the 
one-dimensional Bose gas defined in \eqref{sec3:eq_omega_beta_eps}. Then every Wigner measure associated 
with $(\omega_{\beta,\varepsilon})_{\varepsilon\in(0,1)}$ satisfies the classical $\beta$-KMS condition given in Definition~\ref{classical_KMS_definition}.
\end{theorem*}
We define our model in detail in Section~\ref{sec3:Bose_gas}. Wigner measures 
are defined precisely in Definition~\ref{def:Wigner_measures}. 

\medskip

\paragraph{\bf Outline of the paper:} 
In Section~\ref{sec3:Bose_gas}, we introduce the one-dimensional Bose gases
considered in this work, both on the torus and in the trapped setting on the
real line. We describe the corresponding quantum Gibbs states, the associated
classical nonlinear Schr\"odinger energy, and the Gibbs measure arising in
the classical-field limit and formulate the corresponding classical KMS
condition.
Section~\ref{sec:main_result} introduces Wigner measures, restates our main
result on the derivation of the classical KMS condition, and discusses its
relation to previous works and the main strategy of the proof. In Section~\ref{sec.2.QKMS}, we recall the
quantum KMS condition and establish a quantitative KMS--Gibbs inequality,
which provides the main idea for passing from the quantum equilibrium
identity to its classical counterpart. 
Sections~\ref{sec:derivation_classical_KMS} and~\ref{proofofsemiclassicallemmas} are devoted to the proof of the main theorem: we establish
the required uniform estimates and carry out the semiclassical limit in the
quantum KMS identity, thereby obtaining the classical KMS condition.
Finally, Appendices~\ref{appendix:classical_interaction_energy} and  \ref{Appendix:bounds} respectively collect  additional  results  on integrability of the classical interaction energy
with respect to the free Gaussian measure and some useful free Gibbs-state estimates.


\section{One-dimensional Bose gases}\label{sec3:Bose_gas}

In this section, we briefly recall the basic framework of one-dimensional
Bose gases and fix the notation used throughout the paper. We consider two
settings: a Bose gas on the one-dimensional torus $\mathbb T$ and a trapped
Bose gas on the real line $\mathbb R$.

\subsection{One-particle setting}
Let
\[
\Lambda=\mathbb T=\mathbb R/(2\pi\mathbb Z)
\qquad\text{or}\qquad
\Lambda=\mathbb R,
\]
and denote by
\[
\mathfrak H:=L^2(\Lambda)
\]
the one-particle Hilbert space. Where there is no confusion, we simply
write $L^2$ for $L^2(\Lambda)$. We introduce the following dense subspace of $\mathfrak H=L^2(\Lambda)$: \begin{equation} \label{eq:dense_test_space} 
\mathscr D := 
\begin{cases} 
\mathscr C^\infty(\mathbb T), & \text{if } \Lambda=\mathbb T,\\[1mm] 
\mathscr C^\infty_{\rm c}(\mathbb R), & \text{if } \Lambda=\mathbb R. 
\end{cases} 
\end{equation}

\medskip
We impose the following assumptions on the one-particle Hamiltonian and on
the interaction potential.

\begin{assumption}[One-particle Hamiltonian]
\label{A_assumption}
Let $-\Delta$ denote the non-negative Laplacian on $L^2(\Lambda)$.
We consider the following two cases:
\begin{enumerate}
    \item if $\Lambda=\mathbb T$, we set
    \[
    T:=-\Delta+1;
    \]
    \item if $\Lambda=\mathbb R$, we set
    \[
    T:=-\Delta+|x|^\sigma,
    \qquad \sigma>2.
    \]
\end{enumerate}
\end{assumption}

\begin{assumption}[Interaction potential]
\label{V_assumption}
The interaction potential
\[
V:\Lambda\longrightarrow\mathbb R
\]
is measurable, real-valued, even, and non-negative. Moreover,
\begin{equation}
\label{eq:hyp_V_cases}
V\in
\begin{cases}
L^2(\mathbb T),
& \text{if } \Lambda=\mathbb T,\\[1mm]
L^2_{\mathrm{unif}}(\mathbb R),
& \text{if } \Lambda=\mathbb R,
\end{cases}
\end{equation}
where
\begin{equation}\label{def_L2_unif}
L^2_{\mathrm{unif}}(\mathbb R)
:=
\left\{
V\in L^2_{\mathrm{loc}}(\mathbb R):
\sup_{a\in\mathbb R}
\int_a^{a+1}|V(x)|^2\,\dd x<\infty
\right\}.
\end{equation}
\end{assumption}
We note that
\[
L^p(\mathbb R)\subset L^2_{\mathrm{unif}}(\mathbb R),
\qquad 2\leq p\leq\infty.
\]
More generally, we have
\[
L^2(\mathbb R)+L^\infty(\mathbb R)
\subsetneq L^2_{\mathrm{unif}}(\mathbb R).
\]
In fact, the space $L^2_{\mathrm{unif}}(\mathbb R)$ allows for unbounded functions which can grow over small sets,  such as
\[
V(x) = \sum_{n\in\mathbb Z\setminus\{0\}}
|n|\,
\mathbf 1_{[n,n+|n|^{-2}]}(x).
\]

 With our first assumption, we have the following properties.
\begin{lemma}
\label{A_properties_lemma}
Suppose that $T$ satisfies Assumption \ref{A_assumption}. Then:
\begin{enumerate}
    \item\label{A_property_1}
    $T$ is strictly positive: there exists $c>0$ such that
    \[
    T\geq c\,1;
    \]

    \item\label{A_property_2}
    $T$ has compact resolvent;

    \item\label{A_property_3}
    if $\Lambda=\mathbb T$, then
    \[
    \Tr(T^{-\gamma})<\infty,
    \qquad
    \gamma>\frac12;
    \]
    while, if $\Lambda=\mathbb R$, then
    \[
    \Tr(T^{-\gamma})<\infty,
    \qquad
    \gamma>\frac12+\frac1\sigma.
    \]
   In particular, in both cases,
    \[
    \Tr(T^{-1})<\infty.
    \]
\end{enumerate}
\end{lemma}

\begin{proof}
If $\Lambda=\mathbb T$, the Fourier basis
\[
e_k(x):=(2\pi)^{-1/2}e^{ikx},
\qquad k\in\mathbb Z,
\]
diagonalizes $T$, with
\[
Te_k=(1+k^2)e_k.
\]
Consequently,
\[
\Tr(T^{-\gamma})
=
\sum_{k\in\mathbb Z}(1+k^2)^{-\gamma}<\infty
\]
if and only if $\gamma>\frac12$. Strict positivity and compactness of the
resolvent are immediate.

\medskip
Suppose now that $\Lambda=\mathbb R$. Since $|x|^\sigma\to+\infty$ as
$|x|\to\infty$, the Schr\"odinger operator
\[
T=-\Delta+|x|^\sigma
\]
is strictly positive and has compact resolvent. Moreover, the asymptotic of eigenvalues distribution 
 for one-dimensional  Schr\"odinger operators with increasing potentials
(see, e.g.,~\cite{deWM}) yields
\begin{equation}\label{sec3:eq_Weyl_law}
N_T(E)
:=
\Tr\bigl(\mathbf 1_{(-\infty,E]}(T)\bigr)
\lesssim
E^{\frac12+\frac1\sigma},
\qquad E\to\infty.
\end{equation}
Let
\[
0<\lambda_1\leq \lambda_2\leq\cdots,
\qquad \lambda_n\to+\infty,
\]
denote the eigenvalues of $T$, counted with multiplicity. Taking $E=\lambda_n$ in \eqref{sec3:eq_Weyl_law} yields
\[
n\leq N_T(\lambda_n)
\lesssim
\lambda_n^{\frac12+\frac1\sigma}.
\]
Hence, it follows that
\[
\Tr(T^{-\gamma})=\sum_{n=1}^{\infty}\lambda_n^{-\gamma}<\infty
\qquad\text{whenever}\qquad
\gamma>\frac12+\frac1\sigma.
\]
Since $\sigma>2$, the latter condition is satisfied for $\gamma=1$.
\end{proof}
\begin{remark}\label{rem_T_not_trace}
 Assumption~\ref{A_assumption} implies that the
Gaussian measure with covariance $(\beta T)^{-1}$ is naturally  concentrated on
$L^2$, see Subsection~\ref{sec:class_energy_Gibbs_mes}.  Furthermore, Assumption~\ref{eq:hyp_V_cases} and the integrability condition \eqref{eq:hyp_V_cases}  ensure the almost-sure finiteness of the classical interaction energy, see Appendix~\ref{appendix:classical_interaction_energy}.
\end{remark}

We denote by
\begin{equation}\label{sec3:eq_not_VV}
\mathbb V:=V(x-y)
\end{equation}
the multiplication operator induced by the interaction potential $V$ on
the two-particle Hilbert space $\mathfrak H^{\otimes 2}$. More precisely, $\mathbb V$ acts as
\[
(\mathbb V\Psi)(x,y)
=
V(x-y)\Psi(x,y),
\qquad
\Psi\in \mathcal D(\mathbb V),
\]
where
\[
\mathcal D(\mathbb V)
:=
\left\{
\Psi\in L^2(\Lambda^2):
V(x-y)\Psi(x,y)\in L^2(\Lambda^2)
\right\}.
\]
We denote by $H^1$ the Sobolev space endowed with the norm 
\[
\|u\|_{H^1} = \sqrt{\langle u,(1-\Delta) u\rangle_\mathfrak H}.
\]
We have the following bounds on the interaction potential.

\begin{lemma}[Boundedness of the weighted interaction]
\label{V_properties_lemma}
Assume that $T$ and $V$ satisfy Assumptions~\ref{A_assumption} and
\ref{V_assumption}, respectively. Then
\begin{equation}
\label{eq:bounded_V_sandwiched}
\left(T^{-\frac12}\otimes 1\right)\mathbb V
\in \mathscr B(\mathfrak H^{\otimes 2}),
\qquad
\left(1\otimes T^{-\frac12}\right)\mathbb V
\in \mathscr B(\mathfrak H^{\otimes 2}).
\end{equation}
\end{lemma}

\begin{proof}
We first prove that
\begin{equation}
\label{eq:Vy_T_half_uniform}
\sup_{y\in\Lambda}
\left\|V(\cdot-y)T^{-1/2}\right\|_{\mathscr B(L^2)}
<\infty.
\end{equation}

We distinguish the two cases.

\medskip
\noindent
\emph{Case $\Lambda=\mathbb T$.}
Since $T=1-\Delta,$  the operator $T^{-1/2}$ is a bounded map from $L^2(\mathbb T)$ to $H^1(\mathbb T)$. By the one-dimensional Sobolev embedding
$
H^1(\mathbb T)\hookrightarrow L^\infty(\mathbb T),
$
we obtain, for every $f\in L^2(\mathbb T)$ and uniformly in $y$, 
\begin{align*}
\left\|V(\cdot-y)T^{-1/2}f\right\|_{L^2}
&\lesssim
\|V\|_{L^2}
\left\|T^{-1/2}f\right\|_{H^1}
=
\|V\|_{L^2}\|f\|_{L^2}.
\end{align*}
Thus, we have
{\begin{equation}
\label{eq:Vy_T_half_torus}
\sup_{y\in\mathbb T}
\left\|V(\cdot-y)T^{-1/2}\right\|_{\mathscr B(L^2)}
\lesssim
\|V\|_{L^2(\mathbb T)}.
\end{equation}}

\medskip
\noindent
\emph{Case $\Lambda=\mathbb R$.}
Let
\[
I_k=[k,k+1],
\qquad k\in\mathbb Z.
\]
By the one-dimensional Sobolev embedding theorem on intervals of unit length,
uniformly with respect to $k$, one has 
\[
\|u\|_{L^\infty(I_k)}^2
\le
C\left(
\|u\|_{L^2(I_k)}^2
+
\|\nabla u\|_{L^2(I_k)}^2
\right).
\]

Summing over $k\in\mathbb Z$ and using that the intervals $I_k$
form a partition of $\mathbb R$, we obtain
{\begin{equation}
\label{eq:local_sobolev_sum}
\sum_{k\in\mathbb Z}
\|u\|_{L^\infty(I_k)}^2
\le
C\|u\|_{H^1(\mathbb R)}^2.
\end{equation}
For $u\in H^1(\mathbb R)$, we therefore have
\begin{align}
\left\|V(\cdot-y)u\right\|_{L^2}^2
&=
\sum_{k\in\mathbb Z}
\int_{I_k}
|V(x-y)|^2|u(x)|^2\,\dd x
\nonumber\\
&\leq
\sum_{k\in\mathbb Z}
\|u\|_{L^\infty(I_k)}^2
\int_{I_k}|V(x-y)|^2\, \dd x
\nonumber\\
&\leq
\|V\|_{L^2_{\rm unif}}^2
\sum_{k\in\mathbb Z}
\|u\|_{L^\infty(I_k)}^2
\nonumber\\
&\lesssim
\|V\|_{L^2_{\rm unif}}^2
\|u\|_{H^1(\mathbb R)}^2.
\label{eq:Vunif_H1_multiplier}
\end{align}}
Moreover, since $T$ has compact resolvent and $\ker T=\{0\}$, there exists
$\lambda_0>0$ such that
\[
T\ge \lambda_0.
\]
Therefore
\[
\|u\|_{H^1}^2
=
\langle u,(1-\Delta) u\rangle_\mathfrak H
\le
C\langle u,Tu\rangle_\mathfrak H,
\qquad
u\in 
\mathcal{D}(T^{1/2}).
\]
Taking $u=T^{-1/2}f$, we obtain
\[
\left\|T^{-1/2}f\right\|_{H^1}
\leq
C\|f\|_{L^2}.
\]
Combining this estimate with \eqref{eq:Vunif_H1_multiplier} yields
\begin{equation}
\label{eq:Vy_T_half_R}
\sup_{y\in\mathbb R}
\left\|V(\cdot-y)T^{-1/2}\right\|_{\mathscr B(L^2)}
\lesssim
\|V\|_{L^2_{\rm unif}(\mathbb R)}.
\end{equation}
Therefore \eqref{eq:Vy_T_half_uniform} holds in both cases.

We now pass to the two-particle space. Identifying
$
\mathfrak H^{\otimes2}
\simeq
L^2\bigl(\Lambda_y;L^2(\Lambda_x)\bigr)
$, for $\Psi\in\mathfrak H^{\otimes2}$,
we have
\[
\left[
\mathbb V\left(T^{-1/2}\otimes1\right)\Psi
\right](\cdot,y)
=
V(\cdot-y)T^{-1/2}\Psi(\cdot,y).
\]
Therefore by \eqref{eq:Vy_T_half_uniform}, one has
\begin{align*}
\left\|
\mathbb V\left(T^{-1/2}\otimes1\right)\Psi
\right\|_{L^2(\Lambda^2)}^2
&=
\int_\Lambda
\left\|
V(\cdot-y)T^{-1/2}\Psi(\cdot,y)
\right\|_{L^2_x}^2\,\dd y
\\
&\lesssim
\int_\Lambda
\|\Psi(\cdot,y)\|_{L^2_x}^2\,\dd y
\\
&=
\|\Psi\|_{L^2(\Lambda^2)}^2.
\end{align*}
Thus
\[
\mathbb V\left(T^{-1/2}\otimes1\right)
\in
\mathscr B(\mathfrak H^{\otimes2}).
\]
Since $\mathbb V$ and $T^{-1/2}$ are self-adjoint, taking adjoints gives
\[
\left(T^{-1/2}\otimes1\right)\mathbb V
=
\left[
\mathbb V\left(T^{-1/2}\otimes1\right)
\right]^*
\in
\mathscr B(\mathfrak H^{\otimes2}).
\]
Finally, since $V$ is even, the interaction $\mathbb V$ is invariant under
the exchange of the two variables. The same argument therefore yields
\[
\left(1\otimes T^{-1/2}\right)\mathbb V
\in
\mathscr B(\mathfrak H^{\otimes2}).
\]
\end{proof}

\subsection{Bosonic Fock space and mean-field Hamiltonian}

For $N\geq1$, we denote by
\[
\mathfrak H^{(N)}
:= \mathfrak H\otimes_{\mathrm s} \dots \otimes_{\mathrm s} \mathfrak H
\]
the $N$-fold symmetric tensor product of $\mathfrak H$. Equivalently,
\[
\mathfrak H^{(N)}
=
\left\{
\Psi\in L^2(\Lambda^N):
\Psi(x_{\pi(1)},\ldots,x_{\pi(N)})
=
\Psi(x_1,\ldots,x_N)
\ \text{for every }\pi\in S_N
\right\}.
\]
We use the convention $\mathfrak H^{(0)}=\mathbb C$. The bosonic Fock space
over $\mathfrak H$ is
\[
\mathcal F
:=
\bigoplus_{N=0}^\infty \mathfrak H^{(N)}.
\]
The usual creation and annihilation operators $a^*(f)$ and $a(f)$,
$f\in\mathfrak H$, satisfy the canonical commutation relations
\[
[a(f),a^*(g)]
=
\langle f,g\rangle_{\mathfrak H}\,1,
\qquad
[a(f),a(g)]
=
[a^*(f),a^*(g)]
=
0.
\]
For a semiclassical parameter $\varepsilon>0$, we introduce the rescaled
operators
\[
a_\varepsilon(f)
:=
\sqrt{\varepsilon}\,a(f),
\qquad
a_\varepsilon^*(f)
:=
\sqrt{\varepsilon}\,a^*(f),
\]
which satisfy
\[
[a_\varepsilon(f),a_\varepsilon^*(g)]
=
\varepsilon\langle f,g\rangle_{\mathfrak H}\,1.
\]
If $B$ is a one-particle operator on $\mathfrak H$, its rescaled second
quantization is
\[
\dd\Gamma_\varepsilon(B)
:=
\varepsilon\,\dd\Gamma(B).
\]
In particular,
\[
\dd\Gamma_\varepsilon(B)
\big|_{\mathfrak H^{(N)}}
=
\varepsilon\sum_{j=1}^N B_j.
\]
The rescaled number operator is therefore
\[
\mathcal N_\varepsilon
:=
\dd\Gamma_\varepsilon(1)
=
\varepsilon\mathcal N
=
\varepsilon\bigoplus_{N=0}^\infty N\,1^{(N)}.
\]
Recalling \eqref{sec3:eq_not_VV},  $\mathbb V$ is a two-particle multiplication operator on $\mathfrak H^{(2)}$. 
Its Wick quantization with respect to the rescaled
creation and annihilation operators is formally given by
\[
\mathbb V^{\mathrm{Wick}}
=
\int_{\Lambda^2}
V(x-y)\,
a_\varepsilon^*(x)a_\varepsilon^*(y)
a_\varepsilon(y)a_\varepsilon(x)
\,\dd x\,\dd y.
\]
Equivalently, on the $N$-particle sector,
\[
\mathbb V^{\mathrm{Wick}}
\big|_{\mathfrak H^{(N)}}
=
\varepsilon^2
\sum_{\substack{i,j=1\\i\neq j}}^N
V(x_i-x_j).
\]
We define the grand-canonical quantum Hamiltonian by
\begin{equation}
\label{quantum_Hamiltonian}
H_\varepsilon
:=
\dd\Gamma_\varepsilon(T)
+
\frac12\mathbb V^{\mathrm{Wick}}
=
:H_{\varepsilon,0}
+
\frac12\mathbb V^{\mathrm{Wick}}.
\end{equation}
Since $V\geq 0$, the interaction is repulsive, and hence $H_\varepsilon$ is bounded from below. Moreover, by Lemma \ref{V_properties_lemma}, particle-number conservation, and the KLMN theorem applied sectorwise, $H_\varepsilon$ defines a non-negative self-adjoint operator satisfying
\begin{equation}
H_\varepsilon\geq H_{\varepsilon,0}.
\end{equation}
The corresponding $N$-body mean-field Hamiltonian is
\begin{equation}
\label{N_Hamiltonian}
H_N
:=
\sum_{j=1}^N T_j
+
\frac1{2N}
\sum_{\substack{i,j=1\\i\neq j}}^N
V(x_i-x_j).
\end{equation}
If
\[
\varepsilon=\frac1N
\]
then the two descriptions agree on the $N$-particle sector in the sense that
\begin{equation}
\label{restricted_N_Hamiltonian}
H_\varepsilon\big|_{\mathfrak H^{(N)}}
=
\varepsilon H_N.
\end{equation}
Thus the limit $\varepsilon\to0$ with
$\varepsilon N\simeq1$ corresponds to the usual mean-field regime.

\subsection{Quantum Gibbs states}

Let $\beta>0$ denote the inverse temperature. The grand-canonical quantum
Gibbs density matrix associated with \eqref{quantum_Hamiltonian} is defined by
\begin{equation}
\label{Gibbs_state_definition}
\rho_{\beta,\varepsilon}
:=
\frac{e^{-\beta H_\varepsilon}}
{\Tr_{\mathcal F}(e^{-\beta H_\varepsilon})}.
\end{equation}
We first verify that the corresponding partition function is finite. Throughout, we denote by $\mathfrak S^p$ the Schatten class of order $p$, for  $1\leq p\leq\infty$.

\begin{proposition}
\label{Gibbs_trace_class_proposition}
Suppose that $T$ and $V$ satisfy  Assumptions \ref{A_assumption} and
\ref{V_assumption}, respectively. Then, for every $\beta>0$ and $\varepsilon>0$,
\[
e^{-\beta H_\varepsilon}\in\mathfrak S^1(\mathcal F).
\]
In particular,
\[
0<
\Tr_{\mathcal F}(e^{-\beta H_\varepsilon}) \leq \Tr_{\mathcal F}
\bigl(e^{-\beta H_{\varepsilon,0}}\bigr)
<\infty,
\]
and the Gibbs density matrix \eqref{Gibbs_state_definition} is well defined.
\end{proposition}

\begin{proof}
We first consider the free Hamiltonian
\[
H_{\varepsilon,0}
=
\dd\Gamma_\varepsilon(T).
\]
Let $(\lambda_j)_{j\geq1}$ denote the eigenvalues of $T$, counted with
multiplicity. Since $T>0$ and $T^{-\gamma}$ is trace class by
Lemma \ref{A_properties_lemma}, we have
\[
\sum_{j\geq1}e^{-\beta\varepsilon\lambda_j}<\infty.
\]
The standard bosonic product formula therefore yields
\begin{equation}\label{sec2:stand_bos_prod_form}
\Tr_{\mathcal F}
\bigl(e^{-\beta\dd\Gamma_\varepsilon(T)}\bigr)
=
\prod_{j\geq1}
\frac{1}{1-e^{-\beta\varepsilon\lambda_j}}
<\infty.
\end{equation}
On the other hand, the positivity of $V$ implies that in the sense of quadratic 
forms on $\mathfrak H^{(n)}$, one has
\[
H_\varepsilon^{(n)}
\geq
H_{\varepsilon,0}^{(n)}.
\]
The min-max principle applied on each sector consequently gives
\begin{equation}
\label{eq:min_max_compa_eignvalues}
\lambda_k^{(n)}(\varepsilon)
\geq
\lambda_{k,0}^{(n)}(\varepsilon),
\qquad k\geq1,
\end{equation}
where $\lambda_k^{(n)}(\varepsilon)$ and
$\lambda_{k,0}^{(n)}(\varepsilon)$ denote the eigenvalues of
$H_\varepsilon^{(n)}$ and $H_{\varepsilon,0}^{(n)}$, respectively,
counted with multiplicity and arranged in non-decreasing order. 
Therefore we have 
\begin{equation}\label{eq:comp_partition_int_free}
\Tr_{\mathcal F}
\bigl(e^{-\beta H_\varepsilon}\bigr)
\leq
\Tr_{\mathcal F}
\bigl(e^{-\beta H_{\varepsilon,0}}\bigr)
<\infty.
\end{equation}
\end{proof}
We shall use throughout the notation
\[
Z_{\varepsilon}^{\beta}
:=
\Tr_{\mathcal F}(e^{-\beta H_\varepsilon}),
\qquad
Z_{\varepsilon,0}^{\beta}
:=
\Tr_{\mathcal F}(e^{-\beta H_{\varepsilon,0}})
\]
to denote the interacting and free quantum partition functions respectively.
\begin{definition}\label{def_Gibbs_density_BH}
The Gibbs density matrix $\rho_{\beta,\varepsilon}$, given in \eqref{Gibbs_state_definition}, defines a quantum state
$\omega_{\beta,\varepsilon}$ on $\mathscr B(\mathcal F)$ by
\begin{equation}\label{sec3:eq_omega_beta_eps}
\omega_{\beta,\varepsilon}(A)
:=
\Tr_{\mathcal F}\left(\rho_{\beta,\varepsilon}A\right),
\qquad
A\in\mathscr B(\mathcal F).
\end{equation}
\end{definition}

Note that by Proposition \ref{sec2:prop_Gibbs_kms} below, the state
$\omega_{\beta,\varepsilon}$ is the unique $\beta$-KMS state associated
with the $C^*$-dynamical system
$(\mathscr B(\mathcal F),\alpha^\varepsilon)$, where the Heisenberg dynamics
$\alpha^\varepsilon\equiv(\alpha_t^\varepsilon)_{t\in\mathbb R}$ is generated by
$H_\varepsilon$ and given by
\begin{equation}\label{sec3}
\alpha_t^\varepsilon(A)
:=
e^{\ii t H_\varepsilon} A e^{-\ii t H_\varepsilon},
\qquad
A\in\mathscr B(\mathcal F).
\end{equation}

\subsection{Classical energy and Gibbs measure}
\label{sec:class_energy_Gibbs_mes}
The classical Hamiltonian corresponding to the quantum mean-field
Hamiltonian \eqref{quantum_Hamiltonian} is the nonlinear Schr\"odinger
(NLS) energy
\begin{equation}
\label{classical_energy}
h(u)
=h_0(u)+h_{\mathrm I}(u),
\end{equation}
where
\begin{equation}\label{sec2:eq_h_0+h_I}
h_0(u)
:=
\langle u,Tu\rangle_{\mathfrak H},
\qquad
h_{\mathrm I}(u)
:=
\frac12
\int_{\Lambda^2}
V(x-y)|u(x)|^2|u(y)|^2\,\dd x\,\dd y.
\end{equation}
The formal Hamiltonian flow generated by $h$ is governed by the nonlinear
Schr\"odinger equation
\begin{equation}
\label{classical_NLS}
\ii\partial_tu
=Tu+(V*|u|^2)u.
\end{equation}
We next introduce the associated Gibbs measure. Under
Assumption~\ref{A_assumption}, the operator $T$ has compact resolvent and
satisfies
\begin{equation}
\label{eq:T_inverse_trace_class}
\Tr_{\mathfrak H}(T^{-1})<\infty.
\end{equation}
Consequently, for every $\beta>0$, there exists a centered Gaussian Borel
probability measure $\mu_{\beta,0}$ on $\mathfrak H$ with covariance
$$
\operatorname{Cov}(\mu_{\beta,0})
=
(\beta T)^{-1}.
$$
More explicitly, let $(e_j)_{j\geq1}$ be an orthonormal eigenbasis of
$T$, with
$$
Te_j=\lambda_j e_j.
$$
Then $\mu_{\beta,0}$ is the law of the random series
\begin{equation}
\label{eq:free_Gaussian_series}
u
=
\frac1{\sqrt{\beta}}
\sum_{j\geq1}
\frac{g_j}{\sqrt{\lambda_j}}\,e_j,
\end{equation}
where $(g_j)_{j\geq1}$ is a family of independent standard complex
Gaussian random variables on a probability space $(\Omega,\mathcal A,\mathbb P)$. Indeed,
$$
\mathbb E\bigl[\|u\|_{\mathfrak H}^2\bigr]
=
\frac1\beta
\Tr_{\mathfrak H}(T^{-1})
<\infty,
$$
so that the series \eqref{eq:free_Gaussian_series} converges in
$L^2(\Omega;\mathfrak H)$ and almost surely in $\mathfrak H$. Under Assumption~\ref{V_assumption}, the interaction energy
$h_{\mathrm I}$ is well defined $\mu_{\beta,0}$-almost surely. More
precisely, Proposition~\ref{prop:classical_interaction_finite} yields
\begin{equation}
\label{eq:interaction_energy_integrable}
\int_{\mathfrak H}
h_{\mathrm I}(u)\,
\dd\mu_{\beta,0}(u)
<\infty.
\end{equation}
We  therefore define the interacting classical Gibbs measure by
\begin{equation}
\label{classical_Gibbs_measure}
\dd\mu_\beta(u)
=\frac1{z^\beta}
e^{-\beta h_{\mathrm I}(u)}
\,\dd\mu_{\beta,0}(u),
\end{equation}
where
\begin{equation}
\label{classical_partition_function}
z^\beta
:=
\int_{\mathfrak H}
e^{-\beta h_{\mathrm I}(u)}
\,\dd\mu_{\beta,0}(u).
\end{equation}
Since $V\geq0$, we have
$$
0<e^{-\beta h_{\mathrm I}(u)}\leq1
\qquad
\text{for }\mu_{\beta,0}\text{-almost every }u.
$$
It follows that
$$
0<z^\beta\leq1,
$$
and hence \eqref{classical_Gibbs_measure} defines a Borel probability
measure on $\mathfrak H$.

\subsection{Classical KMS condition}

We  recall the classical KMS condition associated with the NLS
Hamiltonian $h$ introduced in \eqref{classical_energy}-\eqref{sec2:eq_h_0+h_I}. We regard the complex
Hilbert space $\mathfrak H$ as a real Hilbert space endowed with its canonical
symplectic form
\[
\Omega(u,v)
:=
\operatorname{Im}\langle u,v\rangle_{\mathfrak H}.
\]
Let $(e_j)_{j\geq1}$ be an orthonormal eigenbasis of
$T$, and set
\[
f_j:= \ii e_j.
\]
Then $\{e_j,f_j; j\in\mathbb N\}$ is an orthonormal basis of $\mathfrak H$
viewed as a real Hilbert space, with real scalar product
\[
(u,v)_{\mathfrak H_{\mathbb R}}
:=
\operatorname{Re}\langle u,v\rangle_{\mathfrak H}.
\]

\medskip

A function $F:\mathfrak H\to\mathbb R$ is called a \emph{smooth cylindrical
function} if there exist $n\in\mathbb N$ and
$\varphi\in \mathscr C^\infty(\mathbb R^{2n};\mathbb R)$ such that
\begin{equation}
\label{eq:cylindrical-function}
F(u)
=
\varphi\bigl(q_1(u),\ldots,q_n(u),
p_1(u),\ldots,p_n(u)\bigr),
\end{equation}
where
\[
q_j(u):=( u,e_j)_{\mathfrak H_{\mathbb R}}
,
\qquad
p_j(u):=(u,f_j)_{\mathfrak H_{\mathbb R}}.
\]
We denote by $\mathscr C^\infty_{\rm cyl}(\mathfrak H)$ the algebra of such
functions. Next, we shall use the subalgebra
$\mathscr C^\infty_{{\rm cyl},\rm b}(\mathfrak H)$ of bounded smooth
cylindrical functions, obtained by requiring
\[
\varphi\in \mathscr C^\infty_{\rm b}(\mathbb R^{2n};\mathbb R).
\]
Identifying the differential with the gradient through the real
Hilbert structure of $\mathfrak H$, we have for $F\in \mathscr C^\infty_{{\rm cyl},\rm b}(\mathfrak H)$, 
\begin{equation}
\label{eq:gradient-cylindrical}
\nabla F(u)
=
\sum_{j=1}^n
\left[
\partial_{q_j}\varphi\bigl(q(u),p(u)\bigr)e_j
+
\partial_{p_j}\varphi\bigl(q(u),p(u)\bigr)f_j
\right],
\end{equation}
where $q(u)=(q_1(u),\ldots,q_n(u))$ and $p(u)=(p_1(u),\ldots,p_n(u))$.\\
We introduce a canonical  Poisson structure on  the algebra $\mathscr C^\infty_{{\rm cyl},\rm b}(\mathfrak H)$ as follows.  
For $F,G\in\mathscr C^\infty_{{\rm cyl},\rm b}(\mathfrak H)$, real-valued, their Poisson
bracket is defined by
\begin{equation}
\label{eq:poisson-bracket}
\{F,G\}(u)
:=\Omega\bigl(\nabla F(u),\nabla G(u)\bigr).
\end{equation}
 This bracket is
antisymmetric and satisfies the Leibniz and Jacobi identities. Thus
$\mathscr C^\infty_{{\rm cyl},\rm b}(\mathfrak H)$ is a Poisson algebra.  Following \cite{AS23}, we say that a Borel probability measure $\mu$ on $\mathfrak H$ satisfies the classical $\beta$-KMS condition for the NLS
Hamiltonian $h$ in \eqref{classical_energy}-\eqref{sec2:eq_h_0+h_I}  if for any $F,G\in   \mathscr C^\infty_{{\rm cyl},\rm b}(\mathfrak H)$, we have
\begin{equation}
\label{eq.KMS-G}
\int_{\mathfrak H}\, \{F,G\} \,\dd\mu=\beta \int_{\mathfrak H} \,\{ F, h\} \, G\, \dd\mu\,.
\end{equation}
It is shown in \cite[Theorem~2.7]{AS23} that the above classical $\beta$-KMS condition is equivalent to the following statement, which we adopt as it is better suited to our analysis.

\begin{definition}[Classical KMS condition]
\label{classical_KMS_definition}
Let $\beta>0$ be a given inverse temperature. A Borel probability measure $\mu$ on $\mathfrak H$ is said to
satisfy the classical $\beta$-KMS condition associated with the NLS
Hamiltonian $h$ in \eqref{classical_energy}-\eqref{sec2:eq_h_0+h_I}   if
\begin{equation}
\label{sec3:eq:classical_kms_def}
\int_{\mathfrak H}
\left\{
e^{\sqrt{2}\ii\operatorname{Re}\langle g,u\rangle_{\mathfrak H}},
e^{\sqrt{2}\ii\operatorname{Re}\langle f,u\rangle_{\mathfrak H}}
\right\}
\,\dd\mu(u)
=
\beta
\int_{\mathfrak H}
e^{\sqrt{2}\ii\operatorname{Re}\langle f,u\rangle_{\mathfrak H}}
\left\{
e^{\sqrt{2}\ii\operatorname{Re}\langle g,u\rangle_{\mathfrak H}},
h
\right\}
\,\dd\mu(u),
\end{equation}
for all $f,g\in\mathscr D$.
\end{definition}

\begin{remark}
Let us make the following remarks.
\begin{enumerate}
\item[1.] The classical KMS conditions \eqref{eq.KMS-G} and \eqref{sec3:eq:classical_kms_def} can be interpreted as infinite-dimensional integration-by-parts identities associated with the Hamiltonian vector field generated by $h$, see \cite{AS23}.

\item[2.] The Poisson brackets appearing in \eqref{sec3:eq:classical_kms_def} act on complex-valued functions and are defined by complex-linear extension of the real Poisson structure introduced in \eqref{eq:poisson-bracket}.

\item[3.] The formulations \eqref{eq.KMS-G} and \eqref{sec3:eq:classical_kms_def} are  adaptations to our framework of the $\beta$-KMS conditions given in \cite[Definition~2.3]{AS23} and \cite[Theorem~2.7]{AS23}, respectively.

\end{enumerate}
\end{remark}


\section{Main result}\label{sec:main_result}

In this section, we state our main result and place it in the context of the existing literature. 

\subsection{Wigner measures}

Wigner measures were originally introduced in finite-dimensional phase
spaces (see, e.g.,~\cite{PL}) and were subsequently extended to
infinite-dimensional settings by means of projective techniques; see
\cites{ammari2014meanfieldpropagationinfinite,AN08}.

For $f\in\mathfrak H$, we define the Weyl operator
\begin{equation}
\label{Weyl_operator}
W_{\eps}(f):=\e^{\ii\Phi_{\eps}(f)},
\qquad
\Phi_{\eps}(f)
:=
\frac{1}{\sqrt{2}}
\left(a_\eps(f)+a^*_\eps(f)\right).
\end{equation}
For $f,g\in\mathfrak H$, the Weyl operators satisfy the commutation relations
\begin{align}
W_{\eps}(g)^*W_{\eps}(f)W_{\eps}(g)
&=
\e^{-\ii\eps\operatorname{Im}\langle f,g\rangle_{\mathfrak H}}
W_{\eps}(f),
\label{commutationrelations0}
\\
W_{\eps}(g)W_{\eps}(f)
&=
\e^{-\frac{\ii\eps}{2}\operatorname{Im}\langle g,f\rangle_{\mathfrak H}}
W_{\eps}(g+f).
\label{commutationrelations1}
\end{align}
Let $\mathscr P(\mathfrak H)$ denote the set of Borel probability measures on
$\mathfrak H$. We recall the notion of Wigner measure associated with a family
of quantum states. Let $\mathscr A\subset(0,1)$ be such that $0\in\overline{\mathscr A}$.

\begin{definition}[Wigner measure]\label{def:Wigner_measures}
Let $(\omega_\varepsilon)_{\varepsilon\in\mathscr A}$ be a family of quantum states on the $C^*$-algebra $\mathscr B(\mathcal F)$ of bounded operators over the bosonic Fock space $\mathcal F$. A Borel probability measure
$\nu\in\mathscr P(\mathfrak H)$ is called a \emph{Wigner measure} of
$(\omega_\varepsilon)_{\varepsilon\in\mathscr A}$ if there exists a countable
subset $\mathscr A_0\subset\mathscr A$, with
\[
0\in\overline{\mathscr A_0},
\]
such that, for every $f\in\mathfrak H$,
\begin{equation}
\label{defwig}
\lim_{\substack{\varepsilon\to0\\ \varepsilon\in\mathscr A_0}}
\omega_\varepsilon\left(W_\varepsilon(f)
\right)
=
\int_{\mathfrak H}
\e^{\sqrt{2}\ii\operatorname{Re}
\langle f,u\rangle_{\mathfrak H}}
\,\dd\nu(u).
\end{equation}
Equivalently, along $\mathscr A_0$, the quantum characteristic functions
associated with $\omega_\varepsilon$ converge pointwise to the
characteristic function of $\nu$.
\end{definition}

\subsection{Main statement}

Our main result shows that the equilibrium structure
encoded by the quantum KMS condition in Definition~\ref{def:kms:equiv}
is preserved in the semiclassical limit. More precisely, every Wigner
measure arising from the quantum equilibrium states satisfies the
classical KMS relation \eqref{sec3:eq:classical_kms_def}.

\medskip

\begin{theorem}[Quantum-to-classical KMS limit]
\label{quantum_KMS_derivation_theorem}
Let $T$ and $V$ satisfy Assumptions~\ref{A_assumption} and
\ref{V_assumption}, respectively. 
Let $(\omega_{\beta,\varepsilon})_{\varepsilon\in(0,1)}$ be the family of quantum $\beta$-KMS states of the 
one-dimensional Bose gas defined in \eqref{sec3:eq_omega_beta_eps}. Then every Wigner measure associated 
with $(\omega_{\beta,\varepsilon})_{\varepsilon\in(0,1)}$ satisfies the classical $\beta$-KMS condition of Definition~\ref{classical_KMS_definition}.
\end{theorem}

The restriction to one spatial dimension plays an important role in
our analysis. In this regime, the limiting Gibbs field can be treated
without the renormalizations that typically arise in higher
dimensions. 
We also restrict our analysis to models that admit quantum Gibbs states, rather than general quantum KMS states which need not admit a representation by a Gibbs density matrix.

This allows us to exploit
quantitative estimates inherited from the Gibbs density matrix, which
are crucial in controlling the semiclassical limit of the KMS
identity. These restrictions suggest two natural directions for further work:
to extend the argument to more general quantum KMS states, for which
no explicit Gibbs density matrix need be available, and to treat
higher-dimensional models where renormalization becomes necessary.

More broadly, our result supports the following principle: for 
interacting QFTs and Bose gases, semiclassical measures associated
with quantum KMS states should themselves satisfy the corresponding
classical KMS condition. Theorem
\ref{quantum_KMS_derivation_theorem} establishes this principle for
one-dimensional quantum Gibbs states.

\subsection{Background and previous results}

Our result is closely related to, but conceptually distinct from, the
derivation of nonlinear Gibbs measures from quantum many-body
equilibrium states.

For the classical Hamiltonian \eqref{classical_energy}, the
corresponding equilibrium measure is the Gibbs measure
\begin{equation}
\label{eq:intro-classical-gibbs}
\dd\mu_\beta(u)
=
\frac{1}{z^\beta}
\exp\left(
-\frac{\beta}{2}
\bigl\langle |u|^2,V*|u|^2\bigr\rangle
\right)
\dd\mu_{\beta,0}(u),
\end{equation}
where $\mu_{\beta,0}$ denotes the Gaussian measure determined by the free
Hamiltonian and $z_\beta$ is the associated classical partition
function. A substantial body of work, notably by Lewin, Nam, and Rougerie \cite{LNR15,LNR21}, and by Fröhlich, Knowles, Schlein, and Sohinger \cite{FKSS17,FKSS18,FKSS22}, has established the emergence of such nonlinear Gibbs measures from quantum many-body Gibbs states through variational, perturbative, and functional-integral methods.  
A remarkable recent development is the work of Nam, R.~Zhu, and X.~Zhu \cite{NZZ25}, who rigorously derived the three-dimensional $\Phi^4_3$ measure from interacting quantum Bose gases.  We also refer to \cite{CKRG26,FKSS25,RS23,DR24,Soh22,FR26,LNZ26, NZZ26} and the references
therein for further developments.   It is worth noting that in the aforementioned works, convergence to nonlinear Gibbs measures is established at the level of correlation functions, rather than through the convergence of characteristic functions.

The point of view adopted in the present paper is different. Rather
than identifying the limiting Wigner measure directly through
partition functions, reduced density matrices, or functional
integrals, we pass to the semiclassical limit in the
\emph{equilibrium identity itself}.

Our result does not, by itself, identify the limiting Wigner measure
uniquely with the nonlinear Gibbs measure
\eqref{eq:intro-classical-gibbs}. In this respect, it is weaker than
the available direct derivations of Gibbs measures. Its significance
is of a different nature: it shows that the KMS equilibrium structure
itself is stable through  the semiclassical limit. This formulation is
potentially more flexible, since the KMS condition remains meaningful
in settings where an explicit Gibbs density or a direct construction
of the equilibrium measure is unavailable.  The question of whether the classical KMS condition 
\eqref{sec3:eq:classical_kms_def} uniquely characterizes the Gibbs measure will be addressed in future work.

The derivation of a classical KMS condition from microscopic
quantum equilibrium appears to have received considerably less
attention than the derivation of Gibbs measures themselves. In
\cite{AR21}, the first author and Ratsimanetrimanana established a
quantum-to-classical KMS limit for the Bose--Hubbard model on a finite
graph. In \cite{AFP25},  the first three authors developed a quantitative approach based on coherent
states for the finite-dimensional
Bose--Hubbard setting. 
Since the corresponding one-particle phase space is
finite-dimensional, the semiclassical passage avoids several difficulties related to  
lack of compactness and unboundedness issues that arise for continuum field
theories.

The present work extends this strategy to an infinite-dimensional
bosonic system. The underlying mechanism remains the same: one starts
from the quantum KMS identity and passes to its semiclassical limit.
The continuum setting introduces substantial analytical
difficulties, notably the presence of unbounded operators and
non-compact Wick symbols.

We also mention the recent work \cite{DPV25}, see also \cite{Ven24}, where $C^*$-algebraic
methods are used to derive classical KMS states from quantum KMS
states for lattice spin systems. Together with \cite{AR21} and the
present work, these results indicate that the stability of the KMS
condition under suitable classical or semiclassical limits is a
robust phenomenon across different classes of quantum systems.
 In particular, Subsection \ref{subsec:Lp_non_com_est} and 
Proposition \ref{sec2:prop_abstract_KMS_ineq} support the validity of this principle in a broader mathematical framework.

\subsection{Proof strategy}

We conclude this section by outlining the main steps in the proof of
Theorem~\ref{quantum_KMS_derivation_theorem}. This also highlights the
main difficulties caused by the infinite-dimensional nature of the
problem.

The Gibbs state $\omega_{\beta,\eps}$ satisfies the quantum KMS condition for
the Heisenberg dynamics
\[
\alpha_t^\eps(B)
=
\e^{\ii tH_\eps}B\e^{-\ii tH_\eps}.
\]
Formally applying the KMS identity to the Weyl observables
$W_\eps(f)$ and $W_\eps(g)$ yields
\begin{equation}
\label{eq:intro-kms-weyl}
\omega_{\beta,\varepsilon}\left(
W_\eps(f)
\frac{
\alpha^\eps_{\ii\beta}(W_\eps(g))-W_\eps(g)
}{\ii\eps}
\right)
=
\omega_{\beta,\varepsilon}\left(
\frac{
[W_\eps(g),W_\eps(f)]
}{\ii\eps}
\right).
\end{equation}
The right-hand side is comparatively straightforward. Using the Weyl
commutation relations together with the definition of Wigner
measures, one shows that, along any sequence generating a Wigner
measure $\nu$, it converges to the Poisson bracket of the
corresponding classical cylindrical observables.

The left-hand side contains the main difficulty. A second-order
expansion of the imaginary-time dynamics gives the leading term
\begin{equation}
\label{eq:intro-leading}
\beta\,
\omega_{\beta,\varepsilon}\left(
W_\eps(f)
\frac{[W_\eps(g),H_\eps]}{\ii\eps}
\right),
\end{equation}
together with a second-order remainder. Uniform moment estimates for
the number operator, or more generally for suitable modulated-energy
operators, on the interacting Gibbs state allow us to prove that
this remainder vanishes as $\eps\to0$. 

There is, however, an immediate technical obstruction to carrying out
this expansion directly: the imaginary-time factors
$\e^{t\beta H_\eps},$ 
$ t>0,$
are unbounded. We therefore introduce spectral cutoffs for the
Hamiltonian and first perform the expansion at the level of bounded
operators. The cutoffs are subsequently removed. The positive
exponential factors arising in this procedure can be absorbed into
the Gibbs weight, while suitable powers of the number operator provide
the required uniform domination. This constitutes one of the main
technical ingredients in the reduction of the quantum KMS condition.

It remains to identify the semiclassical limit of
\eqref{eq:intro-leading}. Wick calculus allows us to expand the
commutator with the Hamiltonian into a principal contribution and
higher-order terms carrying explicit positive powers of $\eps$.
Uniform number-operator estimates imply that all higher-order
contributions vanish in the limit. The principal contribution is the
Wick quantization of the Poisson bracket between the corresponding
classical Weyl observable and the nonlinear Hamiltonian $h$.

A second difficulty arises at this stage. The operators associated
with the resulting polynomial symbols may be unbounded and, even when
bounded, need not be compact under our assumptions on the interaction
potential $V$. Consequently, the standard convergence results for
Wigner measures cannot be applied directly.

To overcome this lack of compactness, we introduce a spectral
regularization based on the one-particle Hamiltonian. The regularized
symbols are associated with compact operators and therefore fall
within the standard Wigner-measure framework. The difference between
the original and regularized observables is estimated uniformly by
means of fractional free-energy bounds for the interacting Gibbs
state. The resulting error vanishes as the regularization parameter
is removed.

Combining these steps, we obtain
\[
\lim_{\eps\to0}
\omega_{\beta,\varepsilon}\left(
W_\eps(f)
\frac{[W_\eps(g),H_\eps]}{\ii\eps}
\right)
=\frac{1}{2}
\int_{\mfh}
\e^{\sqrt{2}\ii\operatorname{Re}\langle f,u\rangle_{\mathfrak H}}
\left\{
\e^{\sqrt{2}\ii\operatorname{Re}\langle g,u\rangle  _{\mathfrak H}},
h(u)
\right\}
\,\dd\nu(u),
\]
whereas the right-hand side of
\eqref{eq:intro-kms-weyl} converges to
\[
\frac{1}{2}\int_{\mfh}
\left\{
\e^{\sqrt{2}\ii\operatorname{Re}\langle g,u\rangle_{\mathfrak H}},
\e^{\sqrt{2}\ii\operatorname{Re}\langle f,u\rangle_{\mathfrak H}}
\right\}
\,\dd\nu(u).
\]
Passing to the limit in the quantum KMS identity
\eqref{eq:intro-kms-weyl} therefore yields precisely the classical
KMS relation for $\nu$, and concludes the proof of
Theorem~\ref{quantum_KMS_derivation_theorem}.


\section{Quantum KMS condition in abstract settings}
\label{sec.2.QKMS}
Before proceeding with the proof of Theorem \ref{quantum_KMS_derivation_theorem}, we outline the problem in a more abstract setting. In particular, we are able to prove the Duhamel expansion step of the proof in this abstract setting, which gives us \eqref{eq:intro-leading} to leading order. We consider two different general settings (we refer the reader to \cite{BRI,BRII} for more details in this topic). We will work either with a 
$C^*$-dynamical system $(\A,\alpha)$, where $\A$ is a $C^*$-algebra and
\[
  \alpha:\R\longrightarrow \operatorname{Aut}(\A),\qquad t\longmapsto\alpha_t,
\]
is a strongly continuous one-parameter group of $*$-automorphisms. Its infinitesimal generator in this case is the closed derivation $\delta$ defined on
\[
 \mathcal D(\delta)=\left\{A\in\A:\text{norm-}\lim_{t\to0}\frac{\alpha_t(A)-A}{t}\text{ exists in }\A\right\}
\]
by
\[
 \delta(A)=\text{norm-}\lim_{t\to0}\frac{\alpha_t(A)-A}{t}.
\]
Or, we will work with a $W^*$-dynamical system $(\A,\alpha)$, where $\A$ is a von Neumann algebra and  $\alpha$ is a one-parameter group of normal $*$-automorphisms of  $\A$. In this case, the 
the dynamics is  continuous with respect to the $\sigma$-weak topology and need not be norm continuous, precisely 
\[
t\longmapsto \omega(\alpha_t(A))
\quad\text{is continuous for every }
A\in\mathcal A,\text{ and  } \omega\in\mathcal A_*,
\]
where $\mathcal A_*$ denotes the predual of $\A$. Accordingly, the infinitesimal generator 
$\delta$ of the $W^*$-dynamical system is defined by
\[
\delta(A)=\sigma\text{-weak-}\lim_{t\to0}
\frac{\alpha_t(A)-A}{t}.
\]
and its domain is given by 
\[
\mathcal D(\delta)=
\left\{
A\in\A:
\sigma\text{-weak-}\lim_{t\to0}
\frac{\alpha_t(A)-A}{t}
\text{ exists in }\A
\right\}.
\]
An important example for us, is given by the Heisenberg representation. Let $\mathcal H$ be a separable Hilbert space and let $H=H^*$ be a self-adjoint operator on $\mathcal H$. The corresponding Heisenberg dynamics is
\begin{equation}\label{eq:sec2:Heisenberg}
  \alpha_t(A)=e^{\ii tH}Ae^{-\ii tH},\qquad t\in\R.
\end{equation}
Therefore, when $H$ is unbounded, the Heisenberg representation corresponds to the $W^*$ setting, where $\alpha$ is given by \eqref{eq:sec2:Heisenberg} and $\A=\mathscr B(\mathcal H)$ is the $W^*$ or the von Neumann algebra of bounded operators on $\mathcal H$. The predual in this case is the space of trace-class operators
\[
\A_*=\mathcal B(\mathcal H)_*=\mathfrak S^1(\mathcal H).
\]
Also, one formally has
\begin{equation}\label{sec:KMS:derivation_rep_Heis}
 \delta(A)=\ii[H,A].
\end{equation}
Particularly, when $H$ is unbounded, the above commutator formula need to be interpreted with care; the abstract generator $\delta$ is the appropriate object.

\subsection{Entire analytic observables}
We recall the notion of entire analytic observables, which is the same for both the $C^*$ and $W^*$ settings.
\begin{definition}
An element $A\in\A$ is called \emph{entire analytic} for the dynamics $\alpha$ if the orbit map
\[
 \R\ni t\longmapsto\alpha_t(A)\in\A
\]
admits an entire $\A$-valued extension
\[
 \C\ni z\longmapsto\alpha_z(A)\in\A.
\]
We denote the set of entire analytic elements by $\Aan$.
\end{definition}
Equivalently, $A$ is entire analytic if $A\in\bigcap_{n\geq1}\mathcal D(\delta^n)$   and 
\[
 \sum_{n=0}^{\infty}\frac{R^n}{n!}\norm{\delta^n(A)}<\infty
 \quad\text{for every } R>0. 
\]
In this case, one has
\[
 \alpha_z(A)=\sum_{n=0}^{\infty}\frac{z^n}{n!}\delta^n(A),\qquad z\in\C.
\]
For a dynamics generated by $H$, this corresponds formally to the control of all iterated commutators
\[
 \delta^n(A)=\ii^n\ad_H^n(A).
\]
We have the following results, see \cite{BRI} for more details. 

\begin{lemma}
 Let $(\A,\alpha)$ be a $C^*$-(resp.~$W^*$)-dynamical system, then $\Aan$ is a norm (resp.~$\sigma$-weak) dense,  $\delta$ and $\alpha$-invariant $*$-subalgebra of $\A$. 
\end{lemma}

\begin{lemma}[Gaussian approximation]

In both frameworks, for  $A\in\A$ and $\varepsilon>0$, set
\[
 A_\varepsilon
 =\frac{1}{\sqrt{\pi\varepsilon}}
   \int_{\R}e^{-t^2/\varepsilon}\alpha_t(A)\,\dd t.
\]
Then $A_\varepsilon$ is entire analytic and
\[
\begin{cases}
\|A_\varepsilon-A\|\longrightarrow0,
& C^*\text{-dynamical system},\\[2mm]
A_\varepsilon\longrightarrow A
\quad \sigma\text{-weakly},
& W^*\text{-dynamical system}.
\end{cases}
\]
\end{lemma}

In the Heisenberg representation, we  have the following useful approximation, which follows from the spectral theorem. 
\begin{lemma}[Cutoff approximation]\label{sec:KMS:approx_cutoff}
Consider the dynamics  \eqref{eq:sec2:Heisenberg} in the Heisenberg representation.
 Let $\chi$ be a bounded Borel measurable function on $\mathbb R$ such that 
 \[
0\leq \chi\leq 1,\qquad
\chi(x)=1 \quad \text{for } |x|\leq 1,
\qquad
\chi(x)=0 \quad \text{for } |x|\geq 2.
\]
 For $A\in\A$ and $R>0$, set
\[
 A^R
 =\chi\left(\frac H R\right) A \;\chi\left(\frac H R\right).
\]
Then $A^R\in\Aan$ and $A^R$ converges strongly to $A$ as $R$ goes to infinity. 
\end{lemma}

\subsection{KMS condition on entire analytic elements}

 Recall that a normal state on a von Neumann algebra $\A$ is a state that is continuous with respect to the $\sigma$-weak topology.

\begin{definition}[Quantum KMS condition]\label{def:kms:equiv}
Let $(\A,\alpha)$ be a $C^*$ (resp.~$W^*$)-dynamical system and $\beta>0$. A (resp.~normal) state $\omega$ on $\A$ satisfies the $\beta$-KMS condition if for all $A,B\in\Aan$,
 \begin{equation}\label{eq:analytic-kms}
   \omega\bigl(A\alpha_{\ii\beta}(B)\bigr)=\omega(BA).
 \end{equation}
 Such a state is in both cases called $\beta$-KMS state and usually denoted by $\omega_\beta$.  
\end{definition}
For a general $W^*$- or $C^*$-dynamical system $(\A,\alpha)$, a $\beta$-KMS state need not exist for a given $\beta>0$, and when KMS states do exist, they need not be unique. Non-uniqueness is especially common in systems exhibiting phase transitions.
However, in the Heisenberg representation, we do have the following result, see \cite{BRII}.

\begin{proposition}[Gibbs state]\label{sec2:prop_Gibbs_kms}
Assume that $e^{-\beta H}$ is a trace class operator for some $\beta>0$. Then 
\begin{equation}\label{sec:2:eq.Gibbs_state}
\omega_\beta(\cdot) := \frac{\mathrm{Tr}\left(\e^{-\beta H} \, \cdot\right)}{Z^{\beta}}, \qquad Z^{\beta}:= \mathrm{Tr}\left(\e^{-\beta H}\right),
\end{equation}
is the unique normal $\beta$-KMS state of the $W^*$-dynamical system $(\mathcal B(\mathcal H),\alpha)$ with $\alpha$ given by 
\eqref{eq:sec2:Heisenberg}.
\end{proposition}

\subsection{Duhamel principle}

A key argument for the derivation of the classical KMS condition is the following Duhamel principle. 

\begin{proposition}[Duhamel principle]\label{sec:KMS:inv_duhamel_prin}
Let $(\A,\alpha)$ be a $W^*$- or $C^*$-dynamical system and let $\omega$ be an
$\alpha$-invariant state on $\A$. For any $A,B\in \A_{an}$,
\begin{equation}\label{sec2:eq.Duhamel}
\omega(A\alpha_{\ii\beta}(B))-\omega(AB)=\ii\beta\omega(A \delta(B))+\beta^2 \int_0^1 \int_0^t \omega \left( \delta\bigg(\alpha_{-\ii t_1\beta}(A)\bigg) \; 
\delta(B) \right) \, \dd t_1 \dd t.
\end{equation}
\end{proposition}
\begin{proof}
Since $A,B\in\A_{\rm an}$, the maps
\[
z\longmapsto \alpha_z(A),
\qquad
z\longmapsto \alpha_z(B)
\]
are entire $\A$-valued functions. In particular, all the expressions below
are well defined. By the fundamental theorem of calculus applied to
$t\mapsto\alpha_{\ii t\beta}(B)$, we have
\[
\alpha_{\ii\beta}(B)-B
=\ii\beta\int_0^1 \alpha_{\ii t\beta}(\delta(B)) \,\dd t.
\]
Hence
\begin{equation}\label{eq:first-duhamel}
\omega(A\alpha_{\ii\beta}(B))-\omega(AB)
=\ii\beta\int_0^1
\omega\left(A\alpha_{\ii t\beta}(\delta(B))\right) \,\dd t.
\end{equation}
Since $\omega$ is $\alpha$-invariant, its invariance extends from real to
complex times on analytic elements. Thus
\[
\omega\left(A\alpha_{\ii t\beta}(\delta(B))\right)
=\omega\left(\alpha_{-\ii t\beta}(A)\delta(B)\right).
\]
Consequently,
\begin{equation}\label{eq:second-duhamel}
\omega(A\alpha_{\ii\beta}(B))-\omega(AB)
=
\ii\beta\int_0^1
\omega\left(\alpha_{-\ii t\beta}(A)\delta(B)\right) \,\dd t.
\end{equation}
Now set
\[
F(t):=\omega\left(\alpha_{-\ii t\beta}(A)\delta(B)\right), \qquad t\in[0,1].
\]
Since $A$ is entire analytic,
\[
F'(t)=-\ii\beta \, \omega\left(\delta\bigl(\alpha_{-\ii t\beta}(A)\bigr)\delta(B)\right).
\]
Therefore,
\[
F(t) = F(0)-\ii\beta\int_0^t \omega\left(\delta\bigl(\alpha_{-\ii t_1\beta}(A)\bigr)\delta(B)\right) \,\dd t_1.
\]
Since
\[
F(0)=\omega(A\delta(B)),
\]
substituting this identity into \eqref{eq:second-duhamel} gives  precisely \eqref{sec2:eq.Duhamel}.
\end{proof}

For $\beta$-KMS states, this yields the following.
\begin{corollary}[KMS--Duhamel identity]\label{sec:KMS:cor_kms_duhamel_prin}
Let $(\A,\alpha)$ be a $W^*$- or $C^*$-dynamical system and let $\omega_\beta$ be a
$\beta$-KMS state. Then for any $A,B\in\A_{\rm an}$,
\begin{equation}\label{sec2:eq.KMS_Duhamel_ineq}
\omega_\beta([B,A]) = \ii\beta\,\omega_\beta(A\delta(B)) +
\beta^2
\int_0^1\int_0^t
\omega_\beta\!\left(
\delta\bigl(\alpha_{-\ii t_1\beta}(A)\bigr)\,
\delta(B)
\right)
\,\dd t_1\,\dd t .
\end{equation}
\end{corollary}

\begin{proof}
Since every $\beta$-KMS state is invariant under the dynamics $\alpha$, we
can apply the Duhamel principle \eqref{sec2:eq.Duhamel} with $\omega=\omega_\beta$. This gives the claimed identity. 
\end{proof}

In the Heisenberg representation, when the Gibbs state \eqref{sec:2:eq.Gibbs_state} is well defined, with density matrix
\[
    \rho_\beta:=\frac{e^{-\beta H}}{Z^\beta} \in \Schatten^1, 
\]
we establish the following useful inequality. Here, we recall that $\Schatten^p$ denotes the $p$-Schatten class. 

\begin{proposition}[KMS--Gibbs inequality]\label{sec2:prop_kms_Gibbs_ineq}
Let $\omega_\beta$ be the $\beta$-KMS Gibbs state given in Proposition \ref{sec2:prop_Gibbs_kms}. Then for any $A,B\in\A_{\rm an}$,
\begin{equation}\label{sec2:ineq_Gibbs_kms}
\left|\omega_\beta([B,A])-\ii\beta\omega_\beta(A \delta(B)) \right| \leq \beta^2 \int_0^1 \int_0^t
\norm{\delta(A)\rho_\beta^{t_1}}_{\Schatten^{1/t_1}}
    \norm{\delta(B)\rho_\beta^{1-t_1}}_{\Schatten^{1/(1-t_1)}}  \; \dd t_1\, \dd t.
\end{equation}
\end{proposition}
\begin{proof}
By the KMS--Duhamel identity \eqref{sec2:eq.KMS_Duhamel_ineq}, we have
\begin{equation}\label{eq:kms-gibbs-start}
\left|\omega_\beta([B,A])-\ii\beta\,\omega_\beta(A\delta(B))\right|
\leq \beta^2\int_0^1\int_0^t
\left|\omega_\beta\!\left(
\delta\bigl(\alpha_{-\ii t_1\beta}(A)\bigr)\delta(B)
\right)\right|\,\dd t_1\,\dd t .
\end{equation}
It therefore remains to estimate the integrand.

\medskip\noindent 
Recall that
\[
\omega_\beta(C)=\Tr(\rho_\beta C),
\]
and $\alpha_t(C)=e^{\ii tH}Ce^{-\ii tH}$. Since $\delta$ commutes with the
dynamics on analytic elements,
\[
\delta(\alpha_z(A))=\alpha_z(\delta(A)).
\]
For $0\leq t_1\leq1$,
\[
\alpha_{-\ii t_1\beta}(\delta(A))
=\rho_\beta^{-t_1}\delta(A)\rho_\beta^{t_1}.
\]
 Hence
\begin{align*}
\omega_\beta\!\left(
\delta(\alpha_{-\ii t_1\beta}(A))\delta(B)
\right)
&=\Tr\!\left(
\rho_\beta^{1-t_1}\delta(A)\rho_\beta^{t_1}\delta(B)
\right)\\
&=\Tr\!\left(
\delta(A)\rho_\beta^{t_1}\delta(B)\rho_\beta^{1-t_1}
\right).
\end{align*}
For $0<t_1<1$, H\"older's inequality for Schatten classes, with
$p=1/t_1$ and $q=1/(1-t_1)$, gives
\[
\left|
\omega_\beta\!\left(
\delta(\alpha_{-\ii t_1\beta}(A))\delta(B)
\right)
\right|
\leq
\norm{\delta(A)\rho_\beta^{t_1}}_{\Schatten^{1/t_1}}
\norm{\delta(B)\rho_\beta^{1-t_1}}_{\Schatten^{1/(1-t_1)}}.
\]
Substituting this estimate into \eqref{eq:kms-gibbs-start} yields the claimed inequality  \eqref{sec2:ineq_Gibbs_kms}.
\end{proof}

\subsection{Noncommutative $L^p$-spaces associated with a KMS state}\label{subsec:Lp_non_com_est}

 In this paragraph, we extend the \emph{KMS-Gibbs inequality}, stated in a specific $W^*$ setting, to the framework of 
$C^*$-dynamical systems. In particular, this step is well suited for considering  quantum systems in infinite volume. The main argument consists in passing to the von Neumann algebra generated by the GNS representation of a KMS state and using Haagerup $L^p$-spaces and noncommutative H\"older inequalities.

Let $\omega_\beta$ be a $\beta$-KMS state for a $C^*$-dynamical system $(\A,\alpha)$, with
$\beta>0$. We denote its GNS representation by
\[
(\pi_\beta,\mathcal H_\beta,\Omega_\beta),
\]
so that
\[
\omega_\beta(A)
=
\langle\Omega_\beta,\pi_\beta(A)\Omega_\beta\rangle,
\qquad A\in\A.
\]
Passing to the weak closure of $\pi_\beta(\A)$, we introduce the von
Neumann algebra
\[
\M_\beta:=\pi_\beta(\A)''.
\]
Recall that, for a KMS state at nonzero inverse temperature, the GNS vector $\Omega_\beta$ is cyclic and separating for the von Neumann algebra $\M_\beta$, see \cite[Corollary 5.3.9]{BRII}.
Moreover, the state $\omega_\beta$ admits a normal extension to $\M_\beta$, which
we continue to denote by $\omega_\beta$ from now on.

\medskip
 Tomita–Takesaki theory, see \cite[Theorem 5.3.10]{BRII}, associates to the faithful normal (extension of) the state $\omega_\beta$  (to the von Neumann algebra $\M_\beta$) its modular automorphism group
\[
\sigma^{\omega_\beta}
=
\bigl(\sigma_s^{\omega_\beta}\bigr)_{s\in\mathbb R}.
\]
On the other hand, the KMS property identifies this modular group with the original dynamics,
up to the rescaling by the inverse temperature,
\begin{equation}\label{sec2:eq_modular_dynamics}
\sigma_s^{\omega_\beta}\bigl(\pi_\beta(A)\bigr)
=
\pi_\beta\bigl(\alpha_{-\beta s}(A)\bigr),
\qquad A\in\A,\quad s\in\mathbb R.
\end{equation}
For analytic elements this identity extends to complex values of the
parameter. In particular,
\[
\pi_\beta\bigl(\alpha_{-\ii t\beta}(A)\bigr)
=
\sigma_{\ii t}^{\omega_\beta}\bigl(\pi_\beta(A)\bigr),
\qquad A\in\A_{\rm an}.
\]
We shall also use the noncommutative $L^p$-spaces associated with the von
Neumann algebra $\M_\beta$. More precisely, we use the Haagerup spaces
\[
L^p(\M_\beta),
\qquad 1\leq p\leq\infty.
\]
They satisfy
\[
L^\infty(\M_\beta)=\M_\beta,
\]
while $L^1(\M_\beta)$
is canonically identified with the predual
$(\M_\beta)_*$.
The corresponding trace functional
\[
\operatorname{tr}_{\mathrm H}:
L^1(\M_\beta)\longrightarrow\mathbb C
\]
is characterized by
\[
\operatorname{tr}_{\mathrm H}(h_\varphi)
=
\varphi(\mathbf 1),
\qquad \varphi\in(\M_\beta)_*,
\]
where $h_\varphi\in L^1(\M_\beta)$
denotes the density associated with
the normal functional $\varphi$.

\smallskip
In particular, the normal state $\omega_\beta$
satisfies
\begin{equation}\label{W_alg:eq_tr_omega}
\omega_\beta(X)
=
\operatorname{tr}_{\mathrm H}(h_\beta X),
\qquad X\in\M_\beta.
\end{equation}
The density $h_\beta \in L^1(\M_\beta)$ also implements
the modular automorphism group.  Indeed, one has
\[
\sigma_s^{\omega_\beta}(X)
=
h_\beta^{\ii s}Xh_\beta^{-\ii s},
\qquad s\in\mathbb R,\quad X\in\M_\beta.
\]
For entire analytic elements,
the identity extends to complex times.
In particular,
\begin{equation}\label{W_alg:mod_gp_rep}
\sigma_{\ii t}^{\omega_\beta}(X)
=
h_\beta^{-t}Xh_\beta^t,
\qquad X\in(\M_\beta)_{\rm an}.
\end{equation}
This identity is to be understood in the algebra of measurable operators
associated with the Haagerup construction.

\smallskip
For $0<t<1$ and $X\in\M_\beta=L^\infty(\M_\beta)$,
we have
\[
h_\beta^t\in L^{1/t}(\M_\beta),
\qquad
Xh_\beta^t\in L^{1/t}(\M_\beta).
\]
At $t=0$, we use the convention
\[
L^{1/0}(\M_\beta)
:=
L^\infty(\M_\beta)=\M_\beta,
\qquad h_\beta^0=\mathbf 1.
\]
The main analytic ingredient is then the
noncommutative H\"older inequality.
For $1\leq p,q,r\leq\infty$ satisfying
\[
\frac1r=\frac1p+\frac1q,
\]
one has
\[
\norm{XY}_{L^r(\M_\beta)}
\leq
\norm{X}_{L^p(\M_\beta)}
\norm{Y}_{L^q(\M_\beta)}.
\]
In particular, for conjugate exponents
$p,q\in[1,\infty]$,
\begin{equation}\label{W_alg:holder_Lp_noncom}
\left|
\operatorname{tr}_{\mathrm H}(XY)
\right|
\leq
\norm{X}_{L^p(\M_\beta)}
\norm{Y}_{L^q(\M_\beta)}.
\end{equation}
We shall also use the cyclicity property
\[
\operatorname{tr}_{\mathrm H}(XY)
=
\operatorname{tr}_{\mathrm H}(YX),
\]
whenever
$X\in L^p(\M_\beta)$ and
$Y\in L^q(\M_\beta)$ are such that
$1/p+1/q=1$.

\medskip 
We can now formulate the abstract counterpart of the KMS--Gibbs
inequality in Proposition \ref{sec2:prop_kms_Gibbs_ineq}.

\begin{proposition}[KMS inequality in noncommutative $L^p$-spaces]
\label{sec2:prop_abstract_KMS_ineq}
Let $(\A,\alpha)$ be a $C^*$-dynamical system and let $\omega_\beta$ be a
$\beta$-KMS state.  Then for every $A,B\in\A_{\rm an}$,
\begin{align}
\left|
\omega_\beta([B,A])
-\ii\beta\,\omega_\beta(A\delta(B))
\right|
&\leq
\beta^2
\int_0^1\int_0^t
\norm{
\pi_\beta(\delta(A))h_\beta^{t_1}
}_{L^{1/t_1}(\M_\beta)}
\nonumber\\
&\hspace{2cm}\times
\norm{
\pi_\beta(\delta(B))h_\beta^{1-t_1}
}_{L^{1/(1-t_1)}(\M_\beta)}
\,\dd t_1\,\dd t .
\label{sec2:eq_abstract_KMS_ineq}
\end{align}
At $t_1=0$ and $t_1=1$, the norms are interpreted using
\[
L^\infty(\M_\beta)=\M_\beta.
\]
\end{proposition}

\begin{proof}
By the KMS--Duhamel inequality \eqref{eq:kms-gibbs-start},
\begin{equation}\label{sec2:eq_abstract_KMS_start}
\left|\omega_\beta([B,A])-\ii\beta\,\omega_\beta(A\delta(B))\right|
\leq
\beta^2\int_0^1\int_0^t
\left|\omega_\beta\!\left(
\delta(\alpha_{-\ii t_1\beta}(A))\delta(B)
\right)\right|\,\dd t_1\,\dd t .
\end{equation}
It remains to estimate the integrand.

\medskip
The KMS condition identifies the
dynamics with the modular dynamics:
\[
\pi_\beta(\alpha_{-\ii t_1\beta}(C))
=
\sigma_{\ii t_1}^{\omega_\beta}(\pi_\beta(C)).
\]
Since $\delta$ commutes with $\alpha_z$ on analytic elements, setting
\[
X=\pi_\beta(\delta(A)),\qquad Y=\pi_\beta(\delta(B)),
\]
we obtain
\[
\pi_\beta\!\left(\delta(\alpha_{-\ii t_1\beta}(A))\right)
=
\sigma_{\ii t_1}^{\omega_\beta}(X).
\]
The representation of the modular group \eqref{W_alg:mod_gp_rep} yields in terms of the density $h_\beta$,
\[
\sigma_s^{\omega_\beta}(X)=h_\beta^{\ii s}Xh_\beta^{-\ii s},
\]
and hence
\[
\sigma_{\ii t_1}^{\omega_\beta}(X)=h_\beta^{-t_1}Xh_\beta^{t_1}.
\]
Using the Haagerup trace $\operatorname{tr}_{\mathrm H}$  gives
\begin{align*}
\omega_\beta\!\left(
\delta(\alpha_{-\ii t_1\beta}(A))\delta(B)
\right)
=
\operatorname{tr}_{\mathrm H}\!\left(h_\beta\,\sigma_{\ii t_1}^{\omega_\beta}(X)Y\right)
=
\operatorname{tr}_{\mathrm H}\!\left(h_\beta^{1-t_1}Xh_\beta^{t_1}Y\right)=
\operatorname{tr}_{\mathrm H}\!\left(Xh_\beta^{t_1}Yh_\beta^{1-t_1}\right).
\end{align*}

For $0<t_1<1$, set
\[
p=\frac1{t_1},\qquad q=\frac1{1-t_1},
\qquad \frac1p+\frac1q=1.
\]
The noncommutative H\"older inequality \eqref{W_alg:holder_Lp_noncom} yields
\[
\left|
\omega_\beta\!\left(
\delta(\alpha_{-\ii t_1\beta}(A))\delta(B)
\right)
\right|
\leq
\norm{Xh_\beta^{t_1}}_{L^{1/t_1}(\M_\beta)}
\norm{Yh_\beta^{1-t_1}}_{L^{1/(1-t_1)}(\M_\beta)}.
\]
Substitution into \eqref{sec2:eq_abstract_KMS_start} proves
\eqref{sec2:eq_abstract_KMS_ineq}.
\end{proof}


\section{Derivation of the classical KMS condition}
\label{sec:derivation_classical_KMS}

In this section we derive the classical KMS condition from the quantum KMS identity. We first collect the elements of Wick calculus and
Wigner-measure theory that will be needed in the proof. We then establish
uniform estimates for the quantum Gibbs states and use them to justify the
passage to the limit in the quantum KMS identity. Finally, the proof of the main Theorem~\ref{quantum_KMS_derivation_theorem} is presented in Subsection~\ref{Sec:proof_main_thm}.  Throughout this section, we assume  Assumptions~\ref{A_assumption} and
\ref{V_assumption}. 
\subsection{Wick calculus and Wigner measures}
\label{Section:Wick}

We begin by recalling the class of polynomial symbols and their Wick
quantization. 

\begin{definition}[Homogeneous polynomial symbols]
For $p,q\in\mathbb N$, we denote by
$\mathcal P_{p,q}(\mathfrak H)$ the space of functions
$b:\mathfrak H\to\mathbb C$ of the form
\[
b(u)
=
\left\langle
u^{\otimes q},
\widetilde b\,u^{\otimes p}
\right\rangle,
\]
where
\[
\widetilde b
\in
\mathscr B\bigl(\mathfrak H^{(p)},\mathfrak H^{(q)}\bigr).
\]
We denote by $\mathcal P_{p,q}^{\infty}(\mathfrak H)$ the subclass for
which $\widetilde b$ is compact.
\end{definition}

For $b\in\mathcal P_{p,q}(\mathfrak H)$, its Wick quantization
$b^{\mathrm{Wick}}$ is the operator initially defined on the finite-particle
space
\[
\Gamma_{\mathrm{fin}}(\mathfrak H)
:=
\bigoplus_{n=0}^{\mathrm{alg}}\mathfrak H^{(n)}
\]
by
\begin{equation}
\label{wickdef}
b^{\mathrm{Wick}}\big|_{\mathfrak H^{(n)}}
=
\mathbf 1_{\{n\geq p\}}\,
\varepsilon^{\frac{p+q}{2}}
\frac{\sqrt{n!(n+q-p)!}}{(n-p)!}\,
\mathcal S_{n-p+q}
\bigl(
\widetilde b\otimes 1^{(n-p)}
\bigr),
\end{equation}
where $\mathcal S_m$ denotes the orthogonal projection onto
$\mathfrak H^{(m)}$.

\begin{remark}
Formally, if $\widetilde b$ admits an integral kernel, the Wick
quantization may be written in terms of the operator-valued distributions
$a_\varepsilon^*(x)$ and $a_\varepsilon(x)$ as
\[
b^{\mathrm{Wick}}
=
\int
\widetilde b(x_1,\ldots,x_q;y_1,\ldots,y_p)
\prod_{j=1}^q a_\varepsilon^*(x_j)
\prod_{k=1}^p a_\varepsilon(y_k)
\,\dd x\,\dd y.
\]
Formula \eqref{wickdef}, however, will be taken as the rigorous definition.
\end{remark}
For $s\geq0$, we set
\begin{equation}
\label{fractional_lift}
H_{\varepsilon,0}^{(s)}
:=
\dd\Gamma_\varepsilon(T^{s}),
\end{equation}
and use the Japanese bracket notation
\[
\langle x\rangle := (1+|x|^2)^{1/2},
\qquad x\in\mathbb R.
\]

  The next lemma describes the number operator boundedness properties of the Wick observables. Although these observables are unbounded, they can be made bounded by applying a suitable negative power of the number operator; see  for example \cite[Lemma 2.5]{AN08} for the proof.
\begin{lemma}[Number operator bounds for Wick observables]
\label{wickbound}
Let $b\in\mathcal P_{p,q}(\mathfrak H)$. Then there exists
$C_b>0$, independent of $\varepsilon\in(0,1)$, such that
\[
\left\|
\langle\mathcal N_\varepsilon\rangle^{-q/2}
b^{\mathrm{Wick}}
\langle\mathcal N_\varepsilon\rangle^{-p/2}
\right\|_{\mathscr B(\mathcal F)}
\leq C_b.
\]
Therefore
\[
b^{\mathrm{Wick}}
\langle\mathcal N_\varepsilon\rangle^{-(p+q)/2},
\qquad
\langle\mathcal N_\varepsilon\rangle^{-(p+q)/2}
b^{\mathrm{Wick}}
\]
extend to bounded operators on $\mathcal F$, uniformly with respect to
$\varepsilon\in(0,1)$.
\end{lemma}

We also state  some semiclassical estimates for Weyl operators; see for instance \cite[Lemmas 4.4 and 4.5]{AIF_2026__76_2_701_0} for details on the  proof.

\begin{lemma}[Weighted bounds for Weyl operators]
\label{Weylestimates}
Let $f\in \mathscr D$, $m>0$ and  $0<s\leq 1/2$.
Then there exists $C>0$, independent of $\varepsilon\in(0,1)$, such that
\[
\left\|
\langle\mathcal N_\varepsilon\rangle^m
W_\varepsilon(f)
\langle\mathcal N_\varepsilon\rangle^{-m}
\right\|_{\mathscr B(\mathcal F)}
\leq C
\]
and
\[
\left\|
(H_{\varepsilon,0}^{(s)}+1)^{-1/2}
W_\varepsilon(f)
(H_{\varepsilon,0}^{(s)}+1)^{1/2}
\right\|_{\mathscr B(\mathcal F)}
\leq C.
\]
\end{lemma}
\medskip
We now recall  some fundamental results for Wigner measures
that will be used in the sequel. The following result is proved in \cite[Theorem 6.2]{AN08}.

\begin{proposition}[Existence of Wigner measures]
\label{exisenceWignermeasure}
Let $(\varrho_\varepsilon)_{\varepsilon\in(0,1)}$ be a family of density matrices on the bosonic Fock space $\mathcal F$ satisfying
\[
\Tr\bigl(\varrho_\varepsilon\mathcal N_\varepsilon^m\bigr)
\leq C_m
\]
uniformly in $\varepsilon$, for some $m>0$. Then every sequence
$\varepsilon_n\to0$ admits a subsequence, still denoted
$\varepsilon_n$, and a Borel probability measure $\mu$ on $\mathfrak H$
such that $\nu$ is a Wigner measure of
$(\varrho_{\varepsilon_n})_n$. Moreover
\begin{equation}\label{sec5:eq_moment_meas}
\int_{\mathfrak H}\|u\|_{\mathfrak H}^{2m}\,\dd\nu(u)<\infty
\end{equation}
whenever the corresponding uniform moment bound is available.
\end{proposition}

\begin{notation}\label{sec5:not_wig_subset}
Given a sequence $(\varepsilon_k)_{k\in\mathbb N}$ converging to zero, we
denote by
\[
\mathcal M(\varrho_{\varepsilon_k},\,k\in\mathbb N)
\]
the set of all Wigner measures obtained as limits along subsequences of
$(\varrho_{\varepsilon_k})_{k\in\mathbb N}$.
\end{notation}

\medskip

We recall also the following convergence result, which is a consequence of  \cite[Corollary~6.14]{AN08} and Lemma~\ref{Weylestimates}.

 \begin{proposition}
\label{convwithcompactsymbol} 
 Let $(\varrho_\varepsilon)_{\varepsilon\in(0,1)}$ be a family of density matrices on the bosonic Fock space $\mathcal F$ such that 
 \[ 
 \mathcal M(\varrho_{\varepsilon_k},\ k\in\mathbb N) = \{\nu\},
 \] 
 for some sequence $(\varepsilon_k)_{k\in\mathbb N}$ converging to zero. 
Moreover assume that, for every $m>0$, there exists a constant $C_m>0$, independent of $\varepsilon$, such that \begin{equation} \label{eq:number_moment_bound_compact} 
 \Tr\left[ \langle\mathcal N_\varepsilon\rangle^m \varrho_\varepsilon \langle\mathcal N_\varepsilon\rangle^m \right] 
 \leq C_m. 
 \end{equation} 
 Then for every $\eta\in\mathfrak H$ and every  
 \[ 
 c \in \bigoplus_{p,q\in\mathbb N}^{\mathrm{alg}} \mathcal P_{p,q}^{\infty}, 
 \] 
 one has 
 \begin{equation} \label{concompactsymbol} 
  \lim_{k\to\infty} \Tr\left(\varrho_{\varepsilon_k} W_\eps(\eta) \,c^{\mathrm{Wick}}\right) = \int_{\mathfrak H} c(u)\,
  \e^{\sqrt{2}\ii\operatorname{Re}
\langle  \eta,u\rangle_{\mathfrak H}}\,\dd\nu(u). 
 \end{equation} 
 \end{proposition}

\subsection{Uniform estimates for the quantum Gibbs states}
\label{subsec:uniform_Gibbs_estimates}

We now collect useful uniform estimates on the quantum Gibbs states. 
\medskip
We denote the corresponding free Gibbs state by
\[
\rho_{\beta,\varepsilon,0}
:=
\frac{\e^{-\beta H_{\varepsilon,0}}}
{Z^\beta_{\varepsilon,0}}.
\]
 We start by  establishing uniform in \(\varepsilon\) bounds on all number moments of the  Gibbs states.

\begin{lemma}[Uniform number moments]
\label{uniform_moments_number_bounds_lemma}
For every $m>0$, there exists a constant $C_m>0$ (increasing in $m$) such that
\begin{equation}
\label{eq:uniform_number_moments}
\sup_{\varepsilon\in(0,1)}
\Tr_{\mathcal F}
\left[
\rho_{\beta,\varepsilon} (\mathcal N_\varepsilon+1)^m
\right]
\leq C_m .
\end{equation}
\end{lemma}

\begin{proof}

We first recall  that the ratio of the free and interacting partition
functions is uniformly bounded. Indeed,  according to \eqref{eq:comp_partition_int_free}, we have
\[
Z^\beta_\varepsilon
\leq
Z^\beta_{\varepsilon,0}.
\]
Conversely, the Bogoliubov convexity inequality, see \cite[Theorem D.3]{BrZa}, gives
\begin{equation}\label{sec5:2:Peierls_Bogoliubov}
\log Z^\beta_\varepsilon
\geq
\log Z^\beta_{\varepsilon,0}
-
\beta\,
\Tr_{\mathcal F}
\left[
\rho_{\beta,\varepsilon,0}
\left(H_\varepsilon-H_{\varepsilon,0}\right)
\right].
\end{equation}
Using Lemma \ref{AppendixA:lem_2}, the free expectation of the interaction is uniformly bounded in
$\varepsilon$, so that the right-hand side of \eqref{sec5:2:Peierls_Bogoliubov} is bounded from below by $\log Z^\beta_{\varepsilon,0}-C_0$ for some constant $C_0>0$. Hence there exists $C>0$, independent from $\eps$, such that
\begin{equation}
\label{eq:partition_ratio_uniform}
1
\leq
\frac{Z^\beta_{\varepsilon,0}}{Z^\beta_\varepsilon}
\leq C,
\qquad
\forall \varepsilon\in(0,1).
\end{equation}
Using the comparison \eqref{eq:min_max_compa_eignvalues}, one  obtains
\begin{align}
\Tr_{\mathcal F}
\left[
(1+\mathcal N_\varepsilon)^m
\rho_{\beta,\varepsilon}
\right]
&= \sum_{n\geq 0} (1+\varepsilon n)^m\;
\Tr_{\mathfrak H^{(n)}}\left[
\frac{e^{-\beta H_\varepsilon^{(n)}}}{Z_\varepsilon^\beta}
\right]
\\
&\leq
\frac{Z^\beta_{\varepsilon,0}}{Z^\beta_\varepsilon}\,
 \sum_{n\geq 0} (1+\varepsilon n)^m\;
\Tr_{\mathfrak H^{(n)}}
\left[\frac{e^{-\beta H_{\varepsilon,0}^{(n)}}}{Z_{\varepsilon,0}^\beta}
\right]
\\
&\leq
\frac{Z^\beta_{\varepsilon,0}}{Z^\beta_\varepsilon}\;
\Tr_{\mathcal F}
\left[
(1+\mathcal N_\varepsilon)^m
\rho_{\beta,\varepsilon,0}
\right].
\label{eq:number_moment_comparison}
\end{align}
By Lemma \ref{AppendixA:lem_1}, the moments of the free Gibbs state in \eqref{eq:number_moment_comparison} are uniformly bounded. Thus, together with \eqref{eq:partition_ratio_uniform}, this proves 
\eqref{eq:uniform_number_moments}.
\end{proof}

We next record a fractional-energy estimate. Introduce
\begin{equation}
\label{eq:def_sstar}
s_*:=
\begin{cases}
\dfrac12,
& \Lambda=\mathbb T,\\[2mm]
\dfrac12-\dfrac1\sigma,
& \Lambda=\mathbb R.
\end{cases}
\end{equation}
Observe that $s_*>0$ in the trapped case since $\sigma>2$. By
Lemma \ref{A_properties_lemma},
\begin{equation}
\label{eq:fractional_trace_condition}
\Tr\bigl(T^{s-1}\bigr)<\infty,
\qquad
0\leq s<s_*.
\end{equation}
 Recall the notation \eqref{fractional_lift}. We have the following bound on the expectation of $H_{\varepsilon,0}^{(s)}$ in the interacting Gibbs state.

\begin{lemma}[Uniform fractional-energy Wick moments]
\label{lem:uniform_Ts_moments}
Let $0\leq s<s_*$, where $s_*$ is defined in \eqref{eq:def_sstar}. Then
\begin{equation}
\label{eq:uniform_Ts_moments}
\sup_{\varepsilon\in(0,1)}
\Tr_{\mathcal F}\left[\rho_{\beta,\varepsilon}\,H_{\varepsilon,0}^{(s)}\right]<\infty.
\end{equation}
\end{lemma}

\begin{proof}
Fix $0\leq s<s_*$. 
Let
\[
c_0:=\inf\sigma(T)>0,
\qquad
t_0:=\frac12c_0^{1-s}.
\]
For $0\leq t\leq t_0$, consider
\[
H_\varepsilon(t)
:=
H_\varepsilon-tH_{\varepsilon,0}^{(s)}
=
\dd\Gamma_\varepsilon(T-tT^s)+\frac12\mathbb V^{\rm Wick},
\]
and
\[
Z_\varepsilon^\beta(t):=
\Tr_{\mathcal F}\left[e^{-\beta H_\varepsilon(t)}\right].
\]
Since $s<1$ and $T\geq c_0$,  we have, for $0\leq t\leq t_0$,
\begin{equation}\label{sec:5:def_T_s_ineq}
T_{t}:=T-tT^s
\geq
(1-tc_0^{s-1})T
\geq\frac12T.
\end{equation}
Therefore, as in Proposition~\ref{Gibbs_trace_class_proposition} and \eqref{eq:comp_partition_int_free}, we have 
\begin{equation}\label{sec:5:2:eq_Z_t_comp}
Z_\varepsilon^\beta(t)
\leq Z_{\varepsilon,0}^\beta(T_{t}):=
\Tr_{\mathcal F}\left[e^{-\beta\dd\Gamma_\varepsilon(T_{t})}\right]
<\infty.
\end{equation}
By the Bogoliubov convexity inequality in \cite[Theorem D.3]{BrZa}, applied to $H_\varepsilon$ and $H_\varepsilon(t_0)$,
we obtain
\begin{equation}
\label{eq:Ts_convexity}
\beta t_0
\Tr_{\mathcal F}
\left[
\rho_{\beta,\varepsilon}
H_{\varepsilon,0}^{(s)}
\right]
\leq
\log Z_\varepsilon^\beta(t_0)
-\log Z_\varepsilon^\beta.
\end{equation}
On the other hand, we have  by \eqref{sec5:2:Peierls_Bogoliubov}
\begin{equation}\label{sec5:2:Peierls_Bogoliubov_bis}
\log Z_\varepsilon^\beta
\geq
\log Z_{\varepsilon,0}^\beta
-\frac \beta 2\,\Tr_{\mathcal F}\left[\rho_{\beta,\varepsilon,0}
\mathbb V^{\rm Wick}\right].
\end{equation}
Therefore \eqref{eq:Ts_convexity} and \eqref{sec:5:2:eq_Z_t_comp}-\eqref{sec5:2:Peierls_Bogoliubov_bis}  yield 
\begin{align}
\beta t_0
\Tr_{\mathcal F}
\left[\rho_{\beta,\varepsilon}H_{\varepsilon,0}^{(s)}\right]
&\leq
\log Z_\varepsilon^\beta(t_0)-\log Z_\varepsilon^\beta
\nonumber\\
&\leq
\log\frac{Z_{\varepsilon,0}^\beta(T_{t_0})}
{Z_{\varepsilon,0}^\beta}
+
\frac \beta 2\,\Tr_{\mathcal F}\left[\rho_{\beta,\varepsilon,0}\mathbb V^{\rm Wick}\right].
\label{eq:Ts_partition_comparison}
\end{align}
The free Gibbs estimate in Lemma~\ref{AppendixA:lem_2},  proves that $\Tr_{\mathcal F}
\left[\rho_{\beta,\varepsilon,0} \mathbb V^{\rm Wick}\right]$ is uniformly bounded. 

It remains to estimate the free partition function ratio on the right-hand side of \eqref{eq:Ts_partition_comparison}. Applying once more the Bogoliubov convexity inequality \cite[Theorem D.3]{BrZa}, this time to the free Hamiltonians $\dd\Gamma_\varepsilon(T)$ and $\dd\Gamma_\varepsilon(T_{t_0})$,
we obtain
\begin{align}
\log
\frac{Z_{\varepsilon,0}^\beta(T_{t_0})}
{Z_{\varepsilon,0}^\beta}
&\leq
\beta t_0\,
\Tr_{\mathcal F}
\left[
\rho_{\beta,\varepsilon,0}^{\,T_{t_0}}
\;\dd\Gamma_\varepsilon(T^s)
\right]
=
\beta t_0\varepsilon\,
\Tr_{\mathfrak H}
\left[
T^s
\,\left(e^{\beta\varepsilon T_{t_0}}-1\right)^{-1}
\right],
\label{eq:Ts_free_PB}
\end{align}
where $\rho_{\beta,\varepsilon,0}^{\,T_{t_0}}$ denotes the free Gibbs
state associated with the one-particle Hamiltonian  $T_{t_0}$. 
The last equality in \eqref{eq:Ts_free_PB} follows by \eqref{eq:def_reduced_density_matrix}--\eqref{eq:free_1pdm}.  

Applying Lemma~\ref{AppendixA:lem_1},
with $T_{t_0}$ instead of $T$, one has 
\begin{align}
\log
\frac{Z_{\varepsilon,0}^\beta(T_{t_0})}
{Z_{\varepsilon,0}^\beta}
&\leq
t_0\Tr_{\mathfrak H}\left[T^sT_{t_0}^{-1}\right].
\end{align}
Since $T_{t_0}$ and $T$ commute, it follows that
$$
T^sT_{t_0}^{-1}
\leq
2T^{s-1}.
$$
Therefore
\begin{equation}
\label{eq:Ts_free_partition}
\log
\frac{Z_{\varepsilon,0}^\beta(T_{t_0})}
{Z_{\varepsilon,0}^\beta}
\leq
2t_0\Tr_{\mathfrak H}\left[T^{s-1}\right].
\end{equation}
Combining \eqref{eq:Ts_partition_comparison}
 and
\eqref{eq:Ts_free_partition}, we obtain
\[
\Tr_{\mathcal F}
\left[\rho_{\beta,\varepsilon}H_{\varepsilon,0}^{(s)}\right]
\leq
\frac{2}{\beta}\Tr_{\mathfrak H}[T^{s-1}]
+
c_0^{s-1}
\Tr_{\mathcal F}
\left[\rho_{\beta,\varepsilon,0} \mathbb V^{\rm Wick}\right].
\]
By the definition of $s_*$ and Lemma \ref{A_properties_lemma},
\[
\Tr_{\mathfrak H}[T^{s-1}]<\infty.
\]
This completes the proof.  
\end{proof}

\begin{remark}
\label{rem:range_s_uniform_Wick}
The restriction $s<s_*$ is natural. Indeed, even for the free Gibbs state,
the expectation of $\dd\Gamma_\varepsilon(T^{s})$ is controlled by
$\Tr(T^{s-1})$. Hence one cannot in general expect
\eqref{eq:uniform_Ts_moments} to hold up to $s=s_*$. The latter restriction is consistent with the fact that the Gibbs measures concentrate on $H^s$ for
\[
s<\frac12
\quad\text{on }\mathbb T,
\qquad
s<\frac12-\frac1\sigma
\quad\text{on }\mathbb R.
\]

\end{remark}

\subsection{Proof of the main Theorem \ref{quantum_KMS_derivation_theorem}}\label{Sec:proof_main_thm}

We now turn to the main step of the argument. The following proposition
shows that, up to an error vanishing in the semiclassical limit, the
imaginary-time increment appearing in the quantum KMS condition can be
replaced by the commutator with the Hamiltonian. This establishes control of the right hand side of \eqref{sec2:ineq_Gibbs_kms} in Proposition \ref{sec2:prop_kms_Gibbs_ineq}.

\begin{proposition}[First-order reduction of the quantum KMS identity]
\label{tricky_proposition_LHS_KMS}
Let $f,g\in \mathscr D$. 
Then
\begin{equation}
\label{KMS_first_order_reduction}
\omega_{\beta,\varepsilon}
\left(
\frac{[W_{\varepsilon}(g),W_{\varepsilon}(f)]}
{\ii\varepsilon}
\right)
=
\beta\,
\omega_{\beta,\varepsilon}
\left(
W_\varepsilon(f)
\frac{[W_\varepsilon(g),H_\varepsilon]}
{\ii\varepsilon}
\right)
+r_\varepsilon(f,g),
\end{equation}
where
\[
|r_\varepsilon(f,g)|
\leq \varepsilon \,C(f,g,\beta).
\]

\end{proposition}

  We delay the proof of Proposition \ref{tricky_proposition_LHS_KMS} to Section \ref{proofofsemiclassicallemmas}. Having established control of the remainder, it remains to pass to the limit as $\varepsilon\to0$ in \eqref{KMS_first_order_reduction} to recover the classical KMS condition from Definition \ref{classical_KMS_definition}. The following two propositions identify this limit with the expectation of the corresponding Poisson brackets. All propositions below are proved in Section \ref{proofofsemiclassicallemmas}.

\begin{proposition}[Semiclassical limit of the Weyl commutator]
\label{easyterminkms}
Let $f,g\in \mathscr D$.  Assume that for some sequence $(\varepsilon_k)_{k\in\mathbb N}$ converging to zero, 
\[
\mathcal M(\rho_{\beta,\varepsilon_k}, k\in\mathbb N)=\{\nu\}.
\]
Then
\[
\lim_{k\to\infty}
\omega_{\beta,\varepsilon_k}
\left(
\frac{[W_{\varepsilon_k}(g),W_{\varepsilon_k}(f)]}
{\ii\varepsilon_k}
\right)
= \frac{1}{2}
\int_{\mathfrak H}
\left\{
\e^{\sqrt{2}\ii\operatorname{Re}\langle g,u\rangle_{\mathfrak H}},
\e^{\sqrt{2}\ii\operatorname{Re}\langle f,u\rangle_{\mathfrak H}}
\right\}\,\dd\nu(u).
\]
\end{proposition}

\begin{proposition}[Semiclassical limit of the Hamiltonian commutator]
\label{convKMSpart2}
Under the same assumptions,
\[
\lim_{k\to\infty}
\omega_{\beta,\varepsilon_k}
\left(
W_{\varepsilon_k}(f)
\frac{[W_{\varepsilon_k}(g),H_{\varepsilon_k}]}
{\ii\varepsilon_k}
\right)
=\frac{1}{2}
\int_{\mathfrak H}
\e^{\sqrt{2}\ii\operatorname{Re}\langle f,u\rangle_{\mathfrak H}}
\left\{
\e^{\sqrt{2}\ii\operatorname{Re}\langle g,u\rangle_{\mathfrak H}},
h(u)
\right\}\, \dd\nu(u).
\]
\end{proposition}

\medskip

\begin{proof}[\bf Proof of the main Theorem~\ref{quantum_KMS_derivation_theorem}]
By Proposition~\ref{sec2:prop_Gibbs_kms} and \eqref{sec3:eq_omega_beta_eps},
the quantum $\beta$-KMS states $\omega_{\beta,\varepsilon}$ of the one-dimensional Bose gas are
represented by the Gibbs density matrices $\rho_{\beta,\varepsilon}$ defined in
\eqref{Gibbs_state_definition}.  Let $\nu$ be an arbitrary Wigner measure of the family
$(\rho_{\beta,\varepsilon})_{\varepsilon\in(0,1)}$ or equivalently of $(\omega_{\beta,\varepsilon})_{\eps\in(0,1)}$. By definition,
there exists a sequence $(\varepsilon_k)_{k\in\mathbb N}$ converging to
zero such that, 
\[
\mathcal M(\rho_{\beta,\varepsilon_k},\,k\in\mathbb N)
=
\{\nu\}.
\]
Note that the existence of Wigner measures for the family
$(\rho_{\beta,\varepsilon})_{\varepsilon\in(0,1)}$ follows from
Proposition~\ref{exisenceWignermeasure} together with the uniform number
moment estimate of
Lemma~\ref{uniform_moments_number_bounds_lemma}.

We take $f,g$ in  $\mathscr D$. For such $f,g$, the quantum KMS condition satisfied by
$\omega_{\beta,\varepsilon_k}$ gives
\begin{equation}
\label{eq:proof_main_quantum_kms}
\omega_{\beta,\varepsilon_k}
\left(
W_{\varepsilon_k}(f)
\frac{
\alpha^{\varepsilon_k}_{\ii\beta}
\bigl(W_{\varepsilon_k}(g)\bigr)
-W_{\varepsilon_k}(g)}
{\ii\varepsilon_k}
\right)
=
\omega_{\beta,\varepsilon_k}
\left(
\frac{
[W_{\varepsilon_k}(g),W_{\varepsilon_k}(f)]
}
{\ii\varepsilon_k}
\right).
\end{equation}
By Proposition~\ref{tricky_proposition_LHS_KMS}, the left-hand side of
\eqref{eq:proof_main_quantum_kms} admits the expansion
\[
\beta\,
\omega_{\beta,\varepsilon_k}
\left(
W_{\varepsilon_k}(f)
\frac{
[W_{\varepsilon_k}(g),H_{\varepsilon_k}]
}
{\ii\varepsilon_k}
\right)
+o(1),
\qquad k\to\infty.
\]
Therefore, when $k\to\infty$, one has
\begin{equation}
\label{eq:proof_main_approx_kms}
\beta\,
\omega_{\beta,\varepsilon_k}
\left(
W_{\varepsilon_k}(f)
\frac{
[W_{\varepsilon_k}(g),H_{\varepsilon_k}]
}
{\ii\varepsilon_k}
\right)
-
\omega_{\beta,\varepsilon_k}
\left(
\frac{
[W_{\varepsilon_k}(g),W_{\varepsilon_k}(f)]
}
{\ii\varepsilon_k}
\right)
\longrightarrow 0.
\end{equation}

We may now pass to the semiclassical limit in the two terms of
\eqref{eq:proof_main_approx_kms}. Applying
Propositions~\ref{easyterminkms} and~\ref{convKMSpart2}, respectively,
we obtain for $f,g\in\mathscr D$
\begin{align}
\label{eq:proof_classical_kms_dense}
&\int_{\mathfrak H}
\left\{
\e^{\sqrt{2}\ii
\operatorname{Re}\langle g,u\rangle_{\mathfrak H}},
\e^{\sqrt{2}\ii
\operatorname{Re}\langle f,u\rangle_{\mathfrak H}}
\right\}
\,\dd\nu(u)
 =
\beta
\int_{\mathfrak H}
\e^{\sqrt{2}\ii
\operatorname{Re}\langle f,u\rangle_{\mathfrak H}}
\left\{
\e^{\sqrt{2}\ii
\operatorname{Re}\langle g,u\rangle_{\mathfrak H}},
h(u)
\right\}
\,\dd\nu(u).
\end{align}
 This is precisely the classical
$\beta$-KMS condition of
Definition~\ref{classical_KMS_definition}. Hence $\nu$ satisfies the
classical $\beta$-KMS condition, and the proof is complete.
\end{proof}
\begin{remark}
Since $\mathscr D$ is dense in $\mathfrak H$, the continuity
estimate provided by Lemma~\ref{V_properties_lemma}, together with the
moment bound \eqref{sec5:eq_moment_meas} satisfied by the Wigner measure $\nu$, allow us to pass to the limit in 
\eqref{eq:proof_classical_kms_dense} and  extend the identity to arbitrary
$f\in\mathfrak H$ and $g\in \mathcal D(T)$.
\end{remark}


\section{Semiclassical Analysis}\label{proofofsemiclassicallemmas}

In this section, we prove Propositions \ref{tricky_proposition_LHS_KMS}, \ref{easyterminkms} and \ref{convKMSpart2}. Here, we assume  Assumptions~\ref{A_assumption} and
\ref{V_assumption}.


\subsection{Semiclassical commutator expansion}

 We begin by expanding the commutator between the Weyl operator and the Hamiltonian in powers of $\varepsilon$, identifying the leading term and establishing uniform bounds on each term.

\begin{lemma}\label{secondtimeexpansion}
Let $g\in \mathscr D$. We have in the sense of quadratic forms on $D\big(\mathcal N_\eps \sqrt{H_{\varepsilon,0}}\big)$,
\begin{equation*}
\begin{aligned}
  \frac{ \left[ W_{\eps}(g),H_{\eps} \right]}{\ii \eps}
= \sum_{j=1} ^4 \eps^{j-1}   W_{\eps}(g)   \left( \mathcal{R}_j(g)\right) ^{\rm Wick},
\end{aligned} 
\end{equation*}
where we define 
\begin{equation}\label{sec6:eq_def_R_j}
\begin{aligned}
&\mathcal{R}_1(g)[u]:= -\ii \sqrt{2} \mathrm{Im} \langle u, Tg \rangle -\ii \sqrt{2} \mathrm{Im} \langle u^{\otimes 2} , \mathbb V g \otimes u \rangle  \in \mathcal{P}_{0,1}+\mathcal{P}_{1,0}+ \mathcal{P}_{1,2}+ \mathcal{P}_{2,1},
\\ &\mathcal{R}_2(g)[u]:= \frac{\ii}{2} \langle g, Tg \rangle  -\frac{\ii}{2} \Real \langle u^{\otimes 2} ,  \mathbb V g^{\otimes 2} \rangle +\ii \langle g \otimes _s u , \mathbb V g \otimes_s u  \rangle \in  \mathcal{P}_{0,0} + \mathcal{P}_{0,2} + \mathcal{P}_{2,0}+\mathcal{P}_{1,1} ,
 \\ &\mathcal{R}_3(g)[u]:= \frac{\ii}{\sqrt{2} } \mathrm{Im}\langle g ^{\otimes 2} , \mathbb V g \otimes u \rangle  \in \mathcal{P}_{1,0}+\mathcal{P}_{0,1},
 \\ &\mathcal{R}_4(g)[u]:= \frac{\ii }{8} \langle g^{\otimes 2} ,  \mathbb V g^{\otimes 2} \rangle \in \mathcal{P}_{0,0}.
\end{aligned}
\end{equation}
 In particular, the main term $\mathcal{R}_1(g)[u]$ is of the form
\begin{align*}
\mathcal{R}_1(g)[u]   = \frac{1}{\sqrt{2}} 
\Big[ \overline{ b_0(u)}- b_0(u) +b_1(u)-b_2(u) \Big] , 
\end{align*}
with 
\begin{align}
&b_0(u):=\langle  u , Tg\rangle= \langle  u , \tilde b_0 \rangle , 
\hspace{.9in} \tilde b_0 \in \mathscr B (\C,\mfh)\label{mainterm0}
\\& b_1 (u):= \langle g \otimes u , \mathbb V u^{\otimes 2}\rangle= \langle u , \tilde b_1 u^{\otimes 2} \rangle ,
\qquad \tilde b_1:= \ ( \langle g|  \otimes 1 ) \mathbb V  \in \mathscr{B}(  \mfh^{(2)} , \mfh),\label{mainterm1}
\\ & b_2(u):=\langle u^{\otimes 2}, \mathbb V g \otimes u \rangle =\langle u^{\otimes 2} , \tilde b_2 u \rangle, \qquad \tilde b_2:= \mathcal{S}_2 \ \mathbb V (1 \otimes |g\rangle     )   \in \mathscr{B}(  \mfh , \mfh^{(2)}).\label{mainterm2}
\end{align}
Moreover, we have  the following uniform estimates
    \begin{align}
       & \left \Vert \Nop^{-\frac{3}{2}}   \left( \mathcal{R}_j(g)\right) ^{\mathrm{Wick}} \right\Vert_{\mathscr{B}(\mathcal{F})} \leq C,\label{expestimate}
    \end{align}
 for $j\in\{1,2,3,4\}$ and  for some  $C>0$ that may depend  on $g$ but independent of $\eps$.
\end{lemma}

\begin{proof}
Recall that 
\[
h(u)
=
\langle u,Tu\rangle
+
\frac12
\langle u^{\otimes2},\mathbb V u^{\otimes2}\rangle
\]
denote the classical symbol of $H_\varepsilon$, so that
\[
H_\varepsilon=h^{\rm Wick}.
\]
First we note that the commutator $[W_\eps(g), H_\eps]$ can  by computed by Wick calculus  and Weyl relations as a quadratic form on $D\big(\mathcal N_\eps \sqrt{H_{\varepsilon,0}}\big)$ as 

\begin{align*}
[W_\eps(g), H_\eps] & = W_\eps (g) \Big( H_\eps - W_\eps (g)^* H_\eps W_\eps (g) \Big)
\\ & =  W_\eps (g) \Big( h(\cdot)- h(\cdot + \frac{\ii \eps}{\sqrt{2}}  g) \Big )^{\rm Wick},
\end{align*} 
see for example \cite[Proposition 2.9]{AN08}. It thus suffices to expand the polynomial as
\begin{align*}
h(u)-h(u + \frac{\ii \eps}{\sqrt{2}}  g)=\ii\eps\sum_{j=1}^4 \eps^{j-1}\mathcal{R}_j(g)[u],
\end{align*}
which yields exactly the symbols $ \mathcal{R}_j(g) $ given in \eqref{sec6:eq_def_R_j}. Here, we have used that  $\mathbb V $ is even. 

It remains to prove \eqref{expestimate} for $j=1,2,3,4$.
For that, we use the fact that  $g\in\mathscr D$ (in particular $T^{1/2}g\in \mathfrak H$), Lemma \ref{V_properties_lemma}, and the number-operator bounds of Lemma \ref{wickbound} and the Weyl weighted bounds of Lemma~\ref{Weylestimates}. 
\end{proof}


\subsection{First-order semiclassical reduction}\label{Sec:Reduction}

In this subsection we prove Proposition \ref{tricky_proposition_LHS_KMS}, which we break into a number of lemmas.   For the remainder, we fix
\begin{equation*}
1_R(x) := \begin{cases}
1 \quad \text{ if } x \in [-R,R], \\
0 \quad \text{ if } x \notin [-R,R],
\end{cases}
\end{equation*}
and for an operator $A$, define
\begin{equation*}
A^R := 1_R(H_\eps) \,A \,1_R(H_{\eps}).
\end{equation*}
We also take $f,g \in \mathscr D$. 

 The lemma below expresses the quantum KMS condition for the truncated Weyl operators as a leading term and an explicit integral remainder.
\begin{lemma}
\label{cut-off_expansion_lemma}
For any $R,R'>0$ and $\eps \in (0,1)$, we have
\begin{multline*}
\omega_{\beta,\eps} \left( \frac{\left[W^{R'}_\eps(g),W^{R}_{\eps}(f)\right]}{\ii \eps}\right) = \beta \omega_{\beta,\eps} \left( W^{R}_{\eps}(f)\frac{\left[ W^{R'}_\eps(g),H_{\eps} \right]}{\ii \eps}\right) \\
+ \beta^2 \int_0^1 \int_0^{t} \dd t \, \dd t_1  \, \omega_{\beta,\eps} \left( \e^{t_1\beta H_{\eps}} \left[H_{\eps},W^{R}_\eps(f)\right] \e^{-t_1\beta H_{\eps}} \frac{\left[ W^{R'}_\eps(g),H_{\eps}\right]
}{\ii \eps}\right).
\end{multline*}
\end{lemma}
\begin{proof}
This is  a consequence of Corollary \ref{sec:KMS:cor_kms_duhamel_prin}  and Lemma \ref{sec:KMS:approx_cutoff}, see also \eqref{sec:KMS:derivation_rep_Heis}. 
\end{proof}
 We first remove the cutoffs in  the commutator between the Weyl operators.
\begin{lemma}
\label{uniform_bound_expansion_1_lemma_A}
We have
\begin{equation}\label{uniform_bound_expansion_1bis}
\lim_{R,R'\to\infty} 
\omega_{\beta,\eps} \left( \frac{\left[W^{R'}_\eps(g), W^{R}_{\eps}(f)\right]}{\ii \eps}\right) = 
\omega_{\beta,\eps} \left( \frac{\left[W_\eps(g),W_{\eps}(f)\right]}{\ii \eps}\right).
\end{equation}
\end{lemma}
\begin{proof}
The convergence in \eqref{uniform_bound_expansion_1bis} follows directly
from the strong convergence  in Lemma~\ref{sec:KMS:approx_cutoff}.  
\end{proof}
 We next remove the cutoffs in the leading term of the expanded quantum KMS condition.
\begin{lemma}
\label{uniform_bound_expansion_1_lemma_B}
We have
\begin{equation}
\label{uniform_bound_expansion_1}
\lim_{R'\to\infty}  \lim_{R\to\infty}\omega_{\beta,\eps} \left( W^{R}_\eps(f)\frac{\left[ W^{R'}_\eps(g),H_{\eps} \right]}{\ii \eps}\right)= 
\omega_{\beta,\eps} \left( W_\eps(f)\frac{\left[ W_\eps(g),H_{\eps} \right]}{\ii \eps}\right).
\end{equation}
\end{lemma}

\begin{proof}
Throughout the proof $\varepsilon\in(0,1)$ is fixed. We first remove the cutoff $R\to\infty$. Indeed, 
by the strong convergence  in Lemma~\ref{sec:KMS:approx_cutoff}, for every fixed
$R'>0$, we may pass to the limit $R\to\infty$ in \eqref{uniform_bound_expansion_1}.
 
We now consider the limit $R'\to\infty$.  Since $1_{R'}(H_\varepsilon)$ commutes with $H_\varepsilon$ and $1_{R'}\big(H_\varepsilon) \sqrt{H_{\varepsilon,0}^{(s)}}$ is bounded, we can use the quadratic-form identity of
Lemma~\ref{secondtimeexpansion} to expand the commutator 
\begin{align}
\frac{\bigl[W^{R'}_\eps(g),H_\varepsilon\bigr]}
     {\ii\varepsilon}
&=
1_{R'}(H_\varepsilon)\,
\frac{\bigl[W_\eps(g),H_\varepsilon\bigr]}
     {\ii\varepsilon}\,
1_{R'}(H_\varepsilon)
\label{eq:cutoff_commutator}
\\
&=
\sum_{j=1}^4 \varepsilon^{j-1}\,
1_{R'}(H_\varepsilon)\,
W_\varepsilon(g)\,
\bigl(\mathcal R_j(g)\bigr)^{\rm Wick}\,
1_{R'}(H_\varepsilon).
\label{eq:cutoff_commutator_expansion_g} \\
&=
1_{R'}(H_\varepsilon) \, \Sigma_g (\mathcal N_\eps+1)^{3/2} \,
1_{R'}(H_\varepsilon) \label{eq:cutoff_commutator_sigma_1},
\end{align}
where  $\Sigma_g$  is given by 
\begin{align}
\Sigma_g= \sum_{j=1}^4 \varepsilon^{j-1}  W_\varepsilon(g)\,
\bigl(\mathcal R_j(g)\bigr)^{\rm Wick}\, \,(\mathcal N_\eps+1)^{-3/2} \in\mathscr B(\mathcal F),
\label{eq:cutoff_commutator_sigma_2}
\end{align} 
which is a bounded operator on the bosonic Fock space $\mathcal F$ by Lemma~\ref{secondtimeexpansion} and Lemma~\ref{Weylestimates}. We can now justify taking the limit $R'\to \infty$. We write
\begin{equation}
\label{uniform_bound_expansion_3}
\Tr\left[\rho_{\beta,\eps}  W_\eps(f) \,\eqref{eq:cutoff_commutator_sigma_1}_{\rm rhs}\right]= 
\Tr\left[\underbrace{\rho_{\beta,\eps} (\mathcal N_\eps+1)^{3/2}}_{{\in\mathfrak S^1}}\,1_{R'}(H_\varepsilon)W_\eps(f) 1_{R'}(H_\varepsilon)\,\Sigma_g\right].
\end{equation}
Notice that the (uniform) Gibbs estimate in Lemma \ref{uniform_moments_number_bounds_lemma}, yields
\begin{equation}
\label{eq:HS_energy}
\rho_{\beta,\varepsilon}
(\mathcal N_\eps+1)^{3/2}\in\mathfrak S^1.
\end{equation}
This allows us to remove the cutoff $R'\to\infty$ on the right-hand side of \eqref{uniform_bound_expansion_3} again by the strong convergence in Lemma \ref{sec:KMS:approx_cutoff}. 
\end{proof}

 We now show that the remainder is of order $\varepsilon$, uniformly in the cutoffs $R$ and $R'$, and therefore vanishes in the semiclassical limit.

\begin{lemma}\label{main_reminder_est_FOA}
We have that uniformly in $R,R'>0$ and $\eps\in(0,1)$,
\begin{equation}
\label{uniform_bound_expansion_2}
\left|\int_0^1 \int_0^{t} \dd t \, \dd t_1  \, \omega_{\beta,\eps} \left( \e^{t_1\beta H_{\eps}} \left[H_{\eps},W^{R}_\eps(f)\right] \e^{-t_1\beta H_{\eps}} \frac{\left[ W^{R'}_\eps(g),H_{\eps}\right]}{\ii \eps}\right)\right| \leq  \eps \,  C(f,g,\beta) . 
\end{equation}
\end{lemma}

\begin{proof}
For $u\in\mathscr D$ and $R>0$, set
\begin{equation}
\label{eq:def-A-eps-R}
\mathcal A_\varepsilon^R(u)
:=
\frac{[W_\varepsilon^R(u),H_\varepsilon]}{\ii\varepsilon}.
\end{equation}
It suffices to prove that
\begin{equation}
\label{eq:uniform-modular-correlation}
\sup_{\substack{R,R'>0\\0<\varepsilon<1}}
\int_0^1\int_0^t
\left|
\omega_{\beta,\varepsilon}
\left(
\e^{t_1\beta H_\varepsilon}
\mathcal A_\varepsilon^{R}(f)
\e^{-t_1\beta H_\varepsilon}
\mathcal A_\varepsilon^{R'}(g)
\right)
\right|
\,\dd t_1\,\dd t
<\infty.
\end{equation}
We first record a consequence of the commutator expansion.
Since $1_{R'}(H_\varepsilon)$ commutes with $H_\varepsilon$,
Lemma~\ref{secondtimeexpansion} gives
\begin{equation}
\label{eq:A-R-expansion-second}
\mathcal A_\varepsilon^{R'}(u)
=
\sum_{j=1}^4
\varepsilon^{j-1}
1_{R'}(H_\varepsilon)
W_\varepsilon(u)
\bigl(\mathcal R_j(u)\bigr)^{\rm Wick}
1_{R'}(H_\varepsilon).
\end{equation}
The Wick estimates \eqref{expestimate} imply that
\begin{equation}
\label{eq:A-number-weighted}
\sup_{\substack{R'>0\\0<\varepsilon<1}}
\left(
\left\|
(\mathcal N_\varepsilon+1)^{-3/2}
\mathcal A_\varepsilon^{R'}(u)
\right\|
+
\left\|
\mathcal A_\varepsilon^{R'}(u)
(\mathcal N_\varepsilon+1)^{-3/2}
\right\|
\right)
\leq C(u).
\end{equation}
Since $\rho_{\beta,\varepsilon}$ commutes with
$\mathcal N_\varepsilon$, cyclicity of the trace yields, 
for
$0<t_1<1$,
\begin{align}
&
\omega_{\beta,\varepsilon}
\left(
\e^{t_1\beta H_\varepsilon}
\mathcal A_\varepsilon^{R}(f)
\e^{-t_1\beta H_\varepsilon}
\mathcal A_\varepsilon^{R'}(g)
\right)
=
\Tr\left[
\rho_{\beta,\varepsilon}^{\,1-t_1}
\mathcal A_\varepsilon^{R}(f)
\rho_{\beta,\varepsilon}^{\,t_1}
\mathcal A_\varepsilon^{R'}(g)
\right].
\label{eq:rho-fractional-representation}
\end{align}
We distinguish two ranges of $t_1$.

\medskip
\noindent
\emph{Case $0<t_1\leq\frac12$.}
Since $\rho_{\beta,\varepsilon}$ commutes with
$\mathcal N_\varepsilon$, cyclicity,  \eqref{eq:A-number-weighted} and H\"older's inequality for Schatten classes yield
\begin{align*}
&
\left|
\Tr\left[
\rho_{\beta,\varepsilon}^{\,1-t_1}
\mathcal A_\varepsilon^{R'}(f)
\rho_{\beta,\varepsilon}^{\,t_1}
\mathcal A_\varepsilon^R(g)
\right]
\right|
\leq
C(f,g)
\left\|
(\mathcal N_\varepsilon+1)^3
\rho_{\beta,\varepsilon}^{\,1-t_1}
\right\|_{\mathfrak S^{1/(1-t_1)}}
\left\|
\rho_{\beta,\varepsilon}^{\,t_1}
\right\|_{\mathfrak S^{1/t_1}}.
\end{align*}
Since $\Tr\rho_{\beta,\varepsilon}=1$,
\[
\left\|
\rho_{\beta,\varepsilon}^{\,t_1}
\right\|_{\mathfrak S^{1/t_1}}
=1,
\]
and
\begin{align*}
\left\|
(\mathcal N_\varepsilon+1)^3
\rho_{\beta,\varepsilon}^{\,1-t_1}
\right\|_{\mathfrak S^{1/(1-t_1)}}
&=
\left(
\Tr\left[
\rho_{\beta,\varepsilon}
(\mathcal N_\varepsilon+1)^{3/(1-t_1)}
\right]
\right)^{1-t_1}.
\end{align*}
Since $\frac{3}{1-t_1}\leq6$, the uniform number-moment estimate of
Lemma~\ref{uniform_moments_number_bounds_lemma}  yields
\begin{equation}
\label{eq:correlation-first-half}
\left|
\Tr\left[
\rho_{\beta,\varepsilon}^{\,1-t_1}
\mathcal A_\varepsilon^{R'}(f)
\rho_{\beta,\varepsilon}^{\,t_1}
\mathcal A_\varepsilon^R(g)
\right]
\right|
\leq C(f,g,\beta).
\end{equation}

\medskip
\noindent
\emph{Case $\frac12\leq t_1<1$.}
A similar  argument based on Schatten H\"older inequality  gives
\begin{align*}
&
\left|
\Tr\left[
\rho_{\beta,\varepsilon}^{\,1-t_1}
\mathcal A_\varepsilon^{R'}(f)
\rho_{\beta,\varepsilon}^{\,t_1}
\mathcal A_\varepsilon^R(g)
\right]
\right|
\leq
C(f,g)
\left\|
\rho_{\beta,\varepsilon}^{\,1-t_1}
\right\|_{\mathfrak S^{1/(1-t_1)}}
\left\|
(\mathcal N_\varepsilon+1)^3
\rho_{\beta,\varepsilon}^{\,t_1}
\right\|_{\mathfrak S^{1/t_1}}.
\end{align*}
The first factor equals $1$, while
\[
\left\|
(\mathcal N_\varepsilon+1)^3
\rho_{\beta,\varepsilon}^{\,t_1}
\right\|_{\mathfrak S^{1/t_1}}
=
\left(
\Tr\left[
\rho_{\beta,\varepsilon}
(\mathcal N_\varepsilon+1)^{3/t_1}
\right]
\right)^{t_1}.
\]
Since $\frac3{t_1}\leq6$, the same number-moment estimate in
Lemma~\ref{uniform_moments_number_bounds_lemma} gives
\begin{equation}
\label{eq:correlation-second-half}
\left|
\Tr\left[
\rho_{\beta,\varepsilon}^{\,1-t_1}
\mathcal A_\varepsilon^{R'}(f)
\rho_{\beta,\varepsilon}^{\,t_1}
\mathcal A_\varepsilon^R(g)
\right]
\right|
\leq C(f,g,\beta).
\end{equation}
Combining
\eqref{eq:correlation-first-half} and
\eqref{eq:correlation-second-half}, we obtain
\[
\sup_{\substack{R,R'>0\\0<\varepsilon<1\\0<t_1<1}}
\left|
\omega_{\beta,\varepsilon}
\left(
\e^{t_1\beta H_\varepsilon}
\mathcal A_\varepsilon^{R'}(f)
\e^{-t_1\beta H_\varepsilon}
\mathcal A_\varepsilon^R(g)
\right)
\right|
\leq
C(f,g,\beta).
\]
Hence \eqref{eq:uniform-modular-correlation} follows immediately.
This proves \eqref{uniform_bound_expansion_2}.
\end{proof}

We are now able to prove Proposition \ref{tricky_proposition_LHS_KMS}.
\begin{proof}[\bf Proof of Proposition \ref{tricky_proposition_LHS_KMS}]
Using the quantum KMS condition in Lemma \ref{cut-off_expansion_lemma}, combined with Lemmas \ref{uniform_bound_expansion_1_lemma_A}, \ref{uniform_bound_expansion_1_lemma_B} and \ref{main_reminder_est_FOA}, yield  the claimed statement. 
\end{proof}


\subsection{Convergence} \label{Compactnessarguments}

In this subsection, we prove Propositions~\ref{easyterminkms} and \ref{convKMSpart2}.

\begin{proof}[\bf Proof of Proposition  \ref{easyterminkms}]

Assume that for some sequence $(\varepsilon_k)_{k\in\mathbb N}$ converging to zero, 
\[
\mathcal M(\rho_{\beta,\varepsilon_k}, k\in\mathbb N)=\{\nu\}.
\]
By the Weyl commutation relations \eqref{commutationrelations0}-\eqref{commutationrelations1}, we have
\begin{align*}
 \omega_{\beta,\varepsilon_k} \left(  \frac{\left[ W_{\eps_k}(g),W_{\eps_k}(f) \right]}{\ii \eps_k}\right)  
 & =  
 \frac{  1-e^{-\ii\eps_k \mathrm{Im} \langle f,g \rangle_{\mathfrak H}}}{\ii\eps_k}   \,\omega_{\beta,\varepsilon_k} \left(    W_{\eps_k}(g)    W_{\eps_k}(f)  \right)
 \\
 &=   \frac{  e^{-\frac{\ii\eps_k}{2} \mathrm{Im} \langle g,f \rangle_{\mathfrak H}}-e^{\frac{\ii\eps_k}{2} \mathrm{Im} \langle g,f \rangle_{\mathfrak H}}}{\ii\eps_k} \, \omega_{\beta,\varepsilon_k}\left( W_{\eps_k}(g+f)      \right).
\end{align*}
Since $ \mathcal{M}(\rho_{\beta,\varepsilon_k}, \ k \in \mathbb{N} )=\{\nu\}$, for any $\eta \in \mfh$, we have 
\begin{align}\label{defwignermeasure}
    \lim_{k\to \infty}  \omega_{\beta,\varepsilon_k} \left(  W_{\eps_k}(\eta) \right) =  \int_{\mfh} \e^{\sqrt{2}\ii   \mathrm{Re}\langle \eta,u\rangle} \dd \nu (u).
\end{align}
Using \eqref{defwignermeasure} with $\eta= f+g$, we get 
\begin{align*}
\lim_{k\to \infty}  \omega_{\beta,\varepsilon_k}\left(  \frac{\left[ W_{\eps_k}(g),W_{\eps_k}(f) \right]}{\ii \eps_k}\right)
&= \lim_{k\to \infty} \frac{  e^{-\frac{\ii\eps_k}{2} \mathrm{Im} \langle g,f \rangle_{\mathfrak H}}-e^{\frac{\ii\eps_k}{2} \mathrm{Im} \langle g,f \rangle_{\mathfrak H}}}{\ii\eps_k} \; \omega_{\beta,\varepsilon_k} \left(   W_{\eps_k}(g+f)      \right)
\\ & = \int_{\mfh}   {-  \mathrm{Im} \langle g,f \rangle_{\mathfrak H} \, \e^{\sqrt{2}\ii  \mathrm{Re}\langle f+g,u\rangle_{\mathfrak H}}} \dd \nu (u)
\\ & = \frac{1}{2}\int_{\mfh}    \left\{\e^{\sqrt{2}\ii  \mathrm{Re}\langle g,u\rangle_{\mathfrak H}}, \e^{\sqrt{2}\ii  \mathrm{Re}\langle f,u\rangle_{\mathfrak H}} \right\}   \dd \nu (u).
\end{align*}
\end{proof}

\medskip
It remains to prove Proposition~\ref{convKMSpart2}. Since the other terms  carry a factor of $\varepsilon$, we need only  to deal with the convergence of the main part involving $\mathcal{R}_1(g)$,   which by direct computation is given by 
\begin{align*}
\mathcal{R}_1(g)[u]=  \frac{ \ii}{\sqrt{2}}   \left \{ \Real \langle g,u  \rangle , h(u) \right\},
\end{align*}
 where we also recall \eqref{mainterm1}-\eqref{mainterm2}.

 We use an approximation scheme implemented through a  compact regularization of the symbols  $b_1$ and $b_2$. 
Let $\chi \in \mathscr{C}_c^\infty(\R) $ be such that $0 \leq \chi \leq 1$ and 
\begin{align*}
\chi(x)= 
\begin{cases}
1 \qquad |x|\leq 1,
\\  0 \qquad |x|\geq 2.
\end{cases}
\end{align*}
For $M\in \N^*$, we define 
$$\chi_M(x):= \chi\left(\frac{x}{M}\right).$$
Then we will consider the regularised symbols 
\begin{align}
& \tilde b_{1,M} := \chi_M(T^{s}) \  \tilde b_1  \  \mathcal{S}_2,\label{tildeb1M}
\\ & \tilde b_{2,M} := \mathcal{S}_2    \, \tilde b_2 \ \chi_M(T^s),\label{tildeb2M}
\end{align}
where we have $\tilde b_{1,M}^* =  \tilde b_{2,M}$.

\begin{lemma}[Compact regularized operator]\label{regularcompactoperator}
For fixed $M > 0$ and $s>0$, the operators $\tilde b_{1,M}\in \mathscr B(\mfh^{(2)},\mfh)$ and $\tilde b_{2,M} \in \mathscr B(\mfh,\mfh^{(2)})$  defined in \eqref{tildeb1M}-\eqref{tildeb2M}   are compact.
\end{lemma} 
\begin{proof}
For fixed $M>0$ and $s>0$, the operator 
\begin{align*}
    \tilde b_{1,M} = \chi_M(T^s)  \,\tilde{b}_1
\end{align*}
is a compact from $ \mfh^{(2)} $ to $\mfh $ since $\tilde{b}_1$ is bounded by Lemma~\ref{secondtimeexpansion} and $\chi_M(T^{s})$ is compact. By considering the adjoint, the property for  $\tilde{b}_{2,M}$ holds true. 
\end{proof} 

We further need to control the approximation scheme, uniformly  in $\eps\in(0,1)$.
We therefore consider the operators given by
\begin{equation*}
\tilde{B}_{1,M} := (1-\chi_M(T^{s})) \,\tilde b_1, \quad {\tilde{B}}_{2,M} := \tilde b_2\,(1-\chi_M(T^{s})).
\end{equation*}
Recall that for $s\geq0$, 
\begin{equation}
H_{\varepsilon,0}^{(s)}
=
\dd\Gamma_\varepsilon(T^{s}).
\end{equation}

\begin{lemma}
\label{decay_B_lemma}
For $s>0$, one has that
\begin{align}
\label{B_bound_1}
\|(H^{(s)}_{\eps,0}+1)^{-\frac{1}{2}}B_{1,M}^{\rm Wick} (\mcN_\eps+1)^{-1}\|_{\mathscr {B}(\mathcal F)} &\lesssim \frac{1}{\sqrt{M}}, \\
\label{B_bound_2}
\|(\mcN_\eps+1)^{-1}B_{2,M}^{\rm Wick}(H^{(s)}_{\eps,0}+1)^{-\frac{1}{2}}\|_{\mathscr {B}(\mathcal F)} &\lesssim \frac{1}{\sqrt{M}}
\end{align}
where the constants are independent of $\eps \in (0,1)$.
\end{lemma}
\begin{proof}
One has
\begin{align*}
\bigl|\langle \Psi, B_{1,M}^{\mathrm{Wick}}\Phi\rangle\bigr|
&=
\left|
\sum_{n\geq 2}
\bigl\langle
\Psi^{(n-1)},
B_{1,M}^{\mathrm{Wick}}\Phi^{(n)}
\bigr\rangle
\right|
\\
&\leq
\sum_{n\geq 2}
\varepsilon^{\frac32}\sqrt{n(n-1)^2}\,
\left|
\left\langle
\Psi^{(n-1)},
\mathcal S_{n-1}
\bigl(\widetilde B_{1,M}\otimes 1^{\otimes(n-2)}\bigr)
\Phi^{(n)}
\right\rangle
\right|.
\end{align*}
Inserting fractional powers of $T$, we can bound the right hand side by 
\begin{align}
&\sum_{n\geq 2}
\sqrt{\varepsilon(n-1)}
\bigl\|
T^{\frac s 2}\otimes 1^{(n-2)}
\Psi^{(n-1)}
\bigr\|_{\mathfrak H^{(n-1)}}
\bigl\|
T^{-\frac s2}
(1-\chi_M(T^s))
\bigr\|_{\mathscr B(\mathfrak H)}
\\
&\qquad\times
\|\widetilde b_{1}\|_{\mathscr B(\mathfrak H^{(2)},\mathfrak H)}\, \varepsilon n\, 
\|\Phi^{(n)}\|.
\end{align}
Recall that $T$ is strictly positive. Functional calculus gives
\[
\bigl\|
T^{-\frac s2}
(1-\chi_M(T^s))
\bigr\|_{\mathscr B(\mathfrak H)}\lesssim \frac{1}{\sqrt{M}}
\]

Now, using Cauchy--Schwarz 
\begin{align*}
\bigl|\langle \Psi, B_{1,M}^{\mathrm{Wick}}\Phi\rangle\bigr|
&\lesssim
\left(\sum_{n\geq 1}
\bigl\langle
\Psi^{(n)}, 
\dd\Gamma_\varepsilon(T^{s}) \Psi^{(n)}
\bigr\rangle
\right)^{1/2}\,
\left(\sum_{n\geq 1}
\bigl\langle
\Phi^{(n)}, 
\mathcal N_\eps^2 \Phi^{(n)}
\bigr\rangle
\right)^{1/2}\\
&\lesssim  \bigl\| (H_{\varepsilon,0}^{(s)})^{1/2} \Psi \bigr\|_{\mathcal F} \; \bigl\| \mathcal N_\eps\Phi\bigr\|_{\mathcal F}.
\end{align*}
This proves the first estimate. By considering the adjoint, we obtain the second estimate.
\end{proof}
We are now able to prove the following convergence result.  
\begin{proposition}[Convergence of the main term]\label{convmainterm}
For $f,g \in \mathscr D$, we have  
\begin{align*}
 \lim_{k \to \infty}\omega_{\beta,\eps_k} \left(W_{\eps_k}(f)  W_{\eps_k}(g)    \left( \mathcal{R}_1(g)\right) ^{\rm Wick}\right)
& = \int_{\mfh} \e^{\sqrt{2}\ii   \mathrm{Re}\langle f+g,u\rangle_{\mathfrak H}}   
\mathcal{R}_1(g)[u]
\, \dd \nu (u).
\end{align*}
\end{proposition}
\begin{proof}
 We write for $i=1,2$,
\begin{align}
\nonumber
  & \left \vert \omega_{\beta,\eps_k} \left(  W_{\eps_k}(f+g)  \ {b}_{i}^{\rm Wick}\right) - \int_{\mfh} \e^{\sqrt{2}\ii   \mathrm{Re}\langle f+g,u\rangle_{\mathfrak H}}  b_i(u)  \, \dd \nu (u)  \right \vert
\\ & \quad \leq  \left \vert  \omega_{\beta,\eps_k} \left(W_{\eps_k}(f+g) \left( {b}_{i}^{\rm Wick} - {b}_{i,M}^{\rm Wick} \right)\right) \right \vert \label{convestimate1}
\\ & \qquad+\left \vert  \omega_{\beta,\eps_k} \left( W_{\eps_k}(f+g) \  {b}_{i,M}^{\rm Wick}\right) - \int_{\mfh} \e^{\sqrt{2}\ii  \mathrm{Re}\langle f+g,u\rangle_{\mathfrak H}}  b_{i,M}(u)  \, \dd \nu (u)  \right \vert\label{convresult1}
\\ &  \qquad +\left \vert \int_{\mfh} \e^{\sqrt{2}\ii   \mathrm{Re}\langle f+g,u\rangle_{\mathfrak H}} \left(  b_{i,M}(u)-  b_{i}(u)\right)  \, \dd \nu (u) \right \vert. \label{dominatedconvresult1} 
\end{align}
For fixed $M$, Lemma \ref{regularcompactoperator} and \cite[Theorem 6.13]{AN08} implies \eqref{convresult1} converges to zero. For \eqref{dominatedconvresult1}, $b_{i,M}(u) \to b_i(u)$ pointwise, when $M\to\infty$. Moreover, respectively by Lemma~\ref{V_properties_lemma}, Lemma  \ref{uniform_moments_number_bounds_lemma}, and Proposition \ref{exisenceWignermeasure}, we have
\begin{align*}
    \left \vert b_{i,M}(u) \right \vert \lesssim \Vert u\Vert_{\mfh}^3 \quad \text{and } \quad \int_\mfh \Vert u \Vert^3_\mfh \,  \dd \nu(u) <\infty.
\end{align*}
It follows that \eqref{dominatedconvresult1} converges to zero as $M \to \infty$. Finally, picking $0<s<s_*$, where $s_*$ is given by \eqref{eq:def_sstar}, one has that
\begin{multline*}
\left \vert   {\rm Tr } \left( \rho_{\beta,\varepsilon_k}  \ W_{\eps_k}(f+g) \left( {b}_{1}^{\rm Wick} - {b}_{1,M}^{\rm Wick} \right)\right) \right \vert \leq \|(\mcN_{\eps_k} + 1)\rho_{\beta,\eps_k} (H_{\eps_k,0}^{(s)} + 1)^{1/2}\|_{\mathfrak S^1} \\
\times \|(H_{\eps_k,0}^{(s)} + 1)^{-1/2}W_{\eps_k}(f+g) (H_{\eps_k,0}^{(s)} + 1)^{1/2}\|_{\mathscr{B}(\mathcal F)} \|(H_{\eps_k,0}^{(s)} + 1)^{-1/2}B^{\rm Wick}_{1,M}(\mathcal N_{\eps_k}+1)^{-1}\|_{\mathscr{B}(\mathcal F)} \lesssim \frac{1}{\sqrt{M}}.
\end{multline*}
Here we have used Lemmas \ref{decay_B_lemma}, \ref{uniform_moments_number_bounds_lemma}, \ref{lem:uniform_Ts_moments}, and \ref{Weylestimates}.  It follows that the right-hand side of \eqref{convestimate1} converges to zero as $M \to \infty$,  which proves the claimed convergence for $b_1$.  One can argue similarly for $b_{2}$, while the contribution of the term $b_0$ is immediate.  This completes the proof.
\end{proof}
We are now able to prove Proposition~\ref{convKMSpart2}.
\begin{proof}[\bf Proof of Proposition \ref{convKMSpart2}]
Using Lemma \ref{secondtimeexpansion}, number-operator bounds \eqref{expestimate} and Gibbs uniform moment estimates in Lemma \ref{uniform_moments_number_bounds_lemma}, we find
\begin{align}
\omega_{\beta,\eps_k}\left(W_{\eps_k}(f) \frac{ \left[ W_{\eps_k}(g),H_{\eps_k} \right]}{\ii \eps_k}\right)
=\omega_{\beta,\eps_k}\left( W_{\eps_k}(f)  W_{\eps_k}(g)    \left( \mathcal{R}_1(g)\right) ^{\rm Wick}\right) +\mc O(\eps_k).
\end{align}
We then conclude by Proposition \ref{convmainterm}
\begin{align*}
\lim_{k\to \infty} \omega_{\beta,\eps_k} \left( W_{\eps_k}(f) \frac{ \left[ W_{\eps_k}(g),H_{\eps_k} \right]}{\ii \eps_k}\right) &= \lim_{k\to \infty} \omega_{\beta,\eps_k} \left(  W_{\eps_k}(f)  W_{\eps_k}(g)    \left( \mathcal{R}_1(g)\right) ^{\rm Wick}\right)
\\ & =  \int_{\mfh} \e^{\sqrt{2}\ii  \mathrm{Re}\langle f+g,u\rangle_{\mathfrak H}}  \mathcal{R}_1(g)[u]  \dd \nu (u)
\\& =\frac 1{\sqrt 2} \int_{\mfh} \e^{\sqrt{2}\ii   \mathrm{Re}\langle f+g,u\rangle_{\mathfrak H}}  \left( \overline{b_0(u)}- b_0(u)+ b_1(u) -b_2(u) \right)  \dd \nu (u)
\\ & =\frac{1}{2}\int_{\mfh} \e^{\sqrt{2}\ii   \mathrm{Re}\langle f,u\rangle_{\mathfrak H}} \left\{\e^{\sqrt{2}\ii   \mathrm{Re}\langle g,u\rangle_{\mathfrak H}}, h(u) \right\}  \, \dd \nu (u).
\end{align*}
\end{proof}

\appendix
\section{Finiteness of the classical interaction energy}
\label{appendix:classical_interaction_energy}

We justify here the integrability of the classical interaction energy
with respect to the free Gaussian measure.

\begin{lemma}[Trace bound for the interaction]
\label{App.A:V_T_tracebound}
Assume Assumptions~\ref{A_assumption} and~\ref{V_assumption}. Set
\[
K:=T^{-1/2}\otimes T^{-1/2}.
\]
Then the non-negative bounded operator  $K\mathbb V K$
is a  trace-class operator on $\mathfrak H^{(2)}$.  In particular,
\begin{equation}
\label{eq:interaction_trace_appendix}
\Tr_{\mathfrak H^{(2)}}[K\mathbb V K]= \Tr_{\mathfrak H^{(2)}}
\left[
\mathbb V
(T^{-1}\otimes T^{-1})
\right]
<\infty.
\end{equation}
\end{lemma}

\begin{proof}
Let $(e_j)_{j\geq1}$ be an orthonormal eigenbasis of $T$, with
$Te_j=\lambda_j e_j$ and $\lambda_j>0$. On the full tensor product,
$(e_j\otimes e_k)_{j,k\geq1}$ is an orthonormal basis. 
 The non-negative operator
$A:=K\mathbb V K$ is bounded by Lemma~\ref{V_properties_lemma}. Its trace, initially allowed to be
infinite, is given by the sum of its diagonal matrix elements:
\begin{align}
\Tr_{\mathfrak H^{\otimes2}} A
&=\sum_{j,k\geq1}\langle e_j\otimes e_k,A \, e_j\otimes e_k\rangle
\nonumber\\
&=\sum_{j,k\geq1}\frac{1}{\lambda_j\lambda_k}
  \int_{\Lambda^2}V(x-y)|e_j(x)|^2|e_k(y)|^2\,\mathrm dx\,\mathrm dy.
\label{eq:interaction_trace_basis}
\end{align}
By Tonelli's theorem, if
\[
r(x):=\sum_{j\geq1}\frac{|e_j(x)|^2}{\lambda_j}
      =T^{-1}(x,x),
\]
where the last equality denotes the diagonal spectral representation of
the resolvent, then
\begin{equation}
\label{eq:interaction_trace_full}
\Tr_{\mathfrak H^{\otimes2}}A
=\int_{\Lambda^2}V(x-y)r(x)r(y)\,\mathrm dx\,\mathrm dy.
\end{equation}
We show that the right-hand side is finite. First let $\Lambda=\mathbb T$ and 
$T=-\Delta+1$, then
\[
r(x)=\frac1{|\mathbb T|}\sum_{n\in\mathbb Z}\frac1{1+n^2}<\infty.
\]
Then, since $V\geq0$ and $V\in L^2(\mathbb T)$, we have
\[
\Tr_{\mathfrak H^{\otimes2}}A=\int_{\mathbb T^2}V(x-y)r(x)r(y)\,\mathrm dx\,\mathrm dy
\lesssim \|V\|_{L^2(\mathbb T)}<\infty.
\]
Now set $\Lambda=\mathbb R$ and $T=-\Delta+|x|^\sigma$. 
By the one-dimensional inequalities for the diagonal Green function, see Chernyavskaya--Shuster~\cite[Corollary 2.2.1]{ChernyavskayaShuster1999}, one has
$$
T^{-1}(x,x)\lesssim d(x),
$$
where $d(x)$ is the unique positive solution of the equation $$d(x)\int_{x-d(x)}^{x+d(x)} |t|^\sigma\,\dd t=2.$$ 
By checking that $d(x)\lesssim \langle x\rangle^{-\sigma/2}$, it follows that
$$
T^{-1}(x,x)
\lesssim \langle x\rangle^{-\sigma/2},
\qquad x\in\mathbb R.
$$
Hence, we obtain 
\begin{align}
\Tr_{\mathfrak H^{\otimes2}}A \lesssim
\int_{\mathbb R^2}
V(x-y)
\langle x\rangle^{-\sigma/2}
\langle y\rangle^{-\sigma/2}
\,\dd x\,\dd y
\nonumber\
&=
\int_{\mathbb R}
V(z)
\left(
\int_{\mathbb R}
\langle y+z\rangle^{-\sigma/2}
\langle y\rangle^{-\sigma/2}
\,\dd y
\right)
\,\dd z.
\label{eq:interaction_convolution}
\end{align}
Since $\sigma>2$, the standard convolution estimate gives
$$
\int_{\mathbb R}
\langle y+z\rangle^{-\sigma/2}
\langle y\rangle^{-\sigma/2}
\,\dd y
\lesssim
\langle z\rangle^{-\sigma/2}.
$$
Finally, decomposing $\mathbb R$ into the intervals
$I_k=[k,k+1],$ $k\in\mathbb Z$, and using
$V\in L^2_{\rm unif}(\mathbb R)$, we find
\begin{align*}
\int_{\mathbb R}V(z)\langle z\rangle^{-\sigma/2}\,\mathrm dz
&\lesssim\sum_{k\in\mathbb Z}\langle k\rangle^{-\sigma/2}
                 \int_{I_k}V(z)\,\mathrm dz\\
&\lesssim\|V\|_{L^2_{\mathrm{unif}}(\mathbb R)}
                 \sum_{k\in\mathbb Z}\langle k\rangle^{-\sigma/2}<\infty,
\end{align*}
where the last series is finite since $\sigma>2$.

Therefore, we have proved that the right-hand side of
\eqref{eq:interaction_trace_full} is finite and $A$ is trace class. Finally, since  $A$ preserves
$\mathfrak H^{(2)}$, we have also
\[
\operatorname{Tr}_{\mathfrak H^{(2)}}(A|_{\mathfrak H^{(2)}})
\leq\operatorname{Tr}_{\mathfrak H^{\otimes2}}A<\infty.
\]
\end{proof}
\begin{proposition}[Finiteness of the classical interaction energy]
\label{prop:classical_interaction_finite}
Let $T$ and $V$ satisfy Assumptions~\ref{A_assumption} and
\ref{V_assumption} respectively, and let $\mu_{\beta,0}$ be the centered Gaussian
measure with covariance
$$
\operatorname{Cov}(\mu_{\beta,0})
=
(\beta T)^{-1}.
$$
Then the interaction energy term $h_I$ given by \eqref{sec2:eq_h_0+h_I} satisfies
\begin{equation}
\label{eq:interaction_energy_integrable_prop}
\int_{\mathfrak H} h_{\mathrm I}(u)\,\dd\mu_{\beta,0}(u)=\frac 1{\beta^2} \,\Tr_{\mathfrak H^{(2)}}
\left[
\mathbb V
(T^{-1}\otimes T^{-1})
\right] <\infty.
\end{equation}
\end{proposition}

\begin{proof}
Let $(e_j)_{j\geq1}$ be an orthonormal basis of eigenvectors of $T$, with $Te_j=\lambda_j e_j$. We will use the probabilistic representation
\eqref{eq:free_Gaussian_series} and the complex Gaussian Wick formula,
\[
\mathbb E[
\overline g_j\,\overline g_k\,g_\ell g_m]
=
\delta_{j\ell}\delta_{km}
+
\delta_{jm}\delta_{k\ell}.
\]
Therefore, we  can write
\begin{align*}
\label{eq:interaction_energy_integrable_proof}
\int_{\mathfrak H} \langle u^{\otimes2},\mathbb V u^{\otimes2}\rangle \,\dd\mu_{\beta,0}(u)
 &= \frac1{\beta^2}
\sum_{j,k\geq1}
\frac1{\lambda_j\lambda_k}
\Big(
\langle e_j\otimes e_k,
\mathbb V(e_j\otimes e_k)\rangle
+
\langle e_j\otimes e_k,
\mathbb V(e_k\otimes e_j)\rangle
\Big).
\end{align*}
On the other hand, 
\begin{align*}
\Tr_{\mathfrak H^{(2)}}
\left[
\mathbb V(T^{-1}\otimes T^{-1})
\right]
&=
\Tr_{\mathfrak H^{\otimes2}}
\left[
\mathcal S_2\mathbb V(T^{-1}\otimes T^{-1})
\right]
\\
&=
\frac12
\sum_{j,k\geq1}
\frac1{\lambda_j\lambda_k}
\Big(
\langle e_j\otimes e_k,
\mathbb V(e_j\otimes e_k)\rangle
+
\langle e_j\otimes e_k,
\mathbb V(e_k\otimes e_j)\rangle
\Big).
\end{align*}
Therefore
\[
\int_{\mathfrak H}
\langle u^{\otimes2},\mathbb V u^{\otimes2}\rangle
\,\dd\mu_{\beta,0}(u)
=
\frac{2}{\beta^2}
\Tr_{\mathfrak H^{(2)}}
\left[
\mathbb V(T^{-1}\otimes T^{-1})
\right].
\]
The latter quantity is finite by
Lemma~\ref{App.A:V_T_tracebound}.
\end{proof}

\appendix
\setcounter{section}{1}
\section{Free Gibbs state estimates}
\label{Appendix:bounds}
For the reader's convenience, we recall here several known  estimates on free Gibbs states, see, for example, \cite{LNR15}.

\medskip\noindent
  Let $\rho$ be a gauge invariant normal state on the bosonic Fock space $\mathcal F$ with arbitrary finite number moments (i.e., $\rho$ commutes with $\mathcal N_\eps$ in the resolvent sense and 
  $\Tr_{\mathcal F}
\left[ \rho \,\mathcal N_\varepsilon^m \right]<\infty$, for any $m\in \mathbb N$).  We define its $m$-particle reduced density matrix, denoted by
$\rho^{(m)}$  as follows. Writing
\[
\rho=\bigoplus_{n\geq 0}\rho_n,
\]
we define
\begin{equation}
\label{eq:rdm_partial_trace}
\rho^{(m)}
:=
\sum_{n\geq m}
\binom{n}{m}
\Tr_{m+1,\ldots,n}(\rho_n),
\end{equation}
where $\Tr_{m+1,\ldots,n}$ denotes the partial trace over
the last $n-m$ particles. Equivalently, $\rho^{(m)}$ is the unique positive
trace-class operator on $\mathfrak H^{(m)}$ characterized by
\begin{equation}
\label{eq:def_reduced_density_matrix}
m!\,\eps^m
\Tr_{\mathfrak H^{(m)}}
\left[
\rho^{(m)}\,\tilde b
\right]
=
\Tr_{\mathcal F}
\left[
\rho\,b^{\rm Wick}
\right],
\end{equation}
for any $\tilde b\in\mathscr B(\mathfrak H^{(m)})$.
The free Gibbs state 
\[
\rho_{\beta,\varepsilon,0}
=
\frac{e^{-\beta H_{\varepsilon,0}}}
{Z_{\varepsilon,0}^{\beta}}
\]
is a gauge invariant
quasi-free state with arbitrary finite number moments. With the normalization used throughout
the paper, its one-particle density matrix is
\begin{equation}
\label{eq:free_1pdm}
\rho_{\beta,\varepsilon,0}^{(1)}
=
\frac{1}{e^{\beta\varepsilon T}-1}.
\end{equation}
More generally, its  $m$-particle reduced density matrix is given by
\begin{equation}
\label{eq:free_reduced_density}
\rho_{\beta,\varepsilon,0}^{(m)}
=
\left[
\frac{1}{\e^{\beta\varepsilon T}-1}
\right]^{\otimes m},
\end{equation}
where the tensor product is restricted to
$\mathfrak H^{(m)}$ (see for example \cite{LNR15} and \cite[Lemma 4.5]{AmNi_II}).

\begin{lemma}\label{AppendixA:lem_1}
Assume  Assumptions~\ref{A_assumption}.  Then for $m\in\mathbb N$,
\begin{equation}\label{AppA:rho_rdm_m_est}
\varepsilon^m
\rho_{\beta,\varepsilon,0}^{(m)}
\leq \frac{1}{\beta^m}
T^{-1}\otimes\cdots\otimes T^{-1},
\end{equation}
and 
\begin{equation}\label{AppA:number_m_est}
\sup_{\varepsilon\in(0,1)}
\Tr_{\mathcal F}
\left[
\rho_{\beta,\varepsilon,0} \,
\mathcal N_\varepsilon^m
\right]
<\infty.
\end{equation}
\end{lemma}
\begin{proof}
 Functional calculus for $T$ gives
\begin{equation}
\label{eq:free_gamma_bound}
0\leq
\varepsilon\rho_{\beta,\varepsilon,0}^{(1)}
=
\frac{\varepsilon}{e^{\beta\varepsilon T}-1}
\leq
\frac{1}{\beta}T^{-1}.
\end{equation}
Since all the operators involved are positive, taking tensor products
preserves the order, and hence
\[
0\leq
\left(\varepsilon\rho_{\beta,\varepsilon,0}^{(1)}\right)^{\otimes m}
\leq
\frac{1}{\beta^m}
\left(T^{-1}\right)^{\otimes m}.
\]
Using \eqref{eq:free_reduced_density}, we conclude that
\begin{equation}
\varepsilon^m
\rho_{\beta,\varepsilon,0}^{(m)}
\leq \frac{1}{\beta^m}
T^{-1}\otimes\cdots\otimes T^{-1},
\end{equation}
uniformly with respect to $\varepsilon\in(0,1)$. This implies \eqref{AppA:number_m_est}.
\end{proof}

\begin{lemma}\label{AppendixA:lem_2}
Assume  Assumptions~\ref{A_assumption} and
\ref{V_assumption}. Then
\begin{equation}
\label{eq:free_interaction_uniform}
\sup_{\varepsilon\in(0,1)}
\Tr_{\mathcal F}
\left[
\rho_{\beta,\varepsilon,0}\,
\mathbb V^{\rm Wick}
\right]
<\infty.
\end{equation}
\end{lemma}
\begin{proof}
According to \eqref{eq:def_reduced_density_matrix}, we have 
\begin{align}
\Tr_{\mathcal F}
\left[
\rho_{\beta,\varepsilon,0}
\mathbb V^{\rm Wick}
\right]
&=  2\eps^2 \,\Tr_{\mathfrak H^{(2)}}
\left[
\rho_{\beta,\varepsilon,0}^{(2)}\, \mathbb V
\right]
\label{eq:free_interaction_wick}
\end{align}
Using Lemma \ref{AppendixA:lem_1} and the positivity of the potential $V$,  it follows that
\begin{equation}
\label{eq:free_interaction_diagonal}
0\leq
\Tr_{\mathcal F}
\left[
\rho_{\beta,\varepsilon,0}
\mathbb V^{\rm Wick}
\right]
\leq
\frac{ 2}{\beta^2}
\Tr_{\mathfrak H^{(2)}}
\left[
\mathbb V
(T^{-1}\otimes T^{-1})
\right]<\infty.
\end{equation}
The last boundedness is a consequence of Lemma~\ref{App.A:V_T_tracebound}.
\end{proof}

\bigskip

\subsection*{Acknowledgements}
All authors acknowledge funding by the ANR-DFG project "Effective Approximation and Dynamics of Many-Body Quantum Systems" (ANR-22-CE92-0013, DFG project number 505496137).  S.F.~ acknowledges the support of the Deutsche Forschungsgemeinschaft (DFG, German Research Foundation) through CRC/TRR 352 and the University of Tübingen. A.R.~ acknowledges the support by the Italian Ministry of University and Research (MUR) through
the PRIN 2022 grant ``OpeN and Effective quantum Systems (ONES)''.

\subsection*{AI Declaration}
LLMs were used to find the suitable space $L^2_{\mathrm{unif}}(\R)$ for the interaction potential. They were also used for general typesetting and grammar. The rest of the paper was prepared by the authors, who take full responsibility for the validity of the mathematics in this manuscript.

\bibliographystyle{plain}
\bibliography{refs}
\end{document}